\documentclass[twocolumn]{aastex63}

\usepackage{natbib}
\usepackage{txfonts}
\usepackage{rotating}

 \submitjournal{ApJS}

\shorttitle{PGCCs in The First Quadrant and The Anti-center Direction region}
\shortauthors{Zhang et al.}

\begin{document}

\title{Planck Galactic Cold Clumps in Two Regions: the First Quadrant and the Anti-Center Direction Region}

%\correspondingauthor{Yuefang Wu}
%\email{ywu@pku.edu.cn}

\author{Chao Zhang}
\affiliation{Department of Astronomy, Yunnan University, and Key Laboratory of Astroparticle Physics of Yunnan Province, Kunming 650091, China, chaozhangchz@163.com, slqin@bao.ac.cn}

\affiliation{Department of Astronomy, School of Physics, Peking University, Beijing, 100871, China, ywu@pku.edu.cn}

\author{Yuefang Wu}
\affiliation{Department of Astronomy, School of Physics, Peking University, Beijing, 100871, China, ywu@pku.edu.cn}
\affiliation{Kavli Institute for Astronomy and Astrophysics, Peking University, Beijing 100871, China}

\author{Xunchuan Liu}
\affiliation{Department of Astronomy, School of Physics, Peking University, Beijing, 100871, China, ywu@pku.edu.cn}
\affiliation{Kavli Institute for Astronomy and Astrophysics, Peking University, Beijing 100871, China}

\author{Sheng-li Qin}
\affiliation{Department of Astronomy, Yunnan University, and Key Laboratory of Astroparticle Physics of Yunnan Province, Kunming 650091, China, chaozhangchz@163.com, slqin@bao.ac.cn}

\author{Tie Liu}
\affiliation{Shanghai Astronomical Observatory, Chinese Academy of Sciences, 80 Nandan Road, Shanghai 200030, China}

\author{Jinghua Yuan}
\affiliation{National Astronomical Observatories, Chinese Academy of Sciences, 20A Datun Road, Chaoyang District, Beijing 100012, China}

\author{Di Li}
\affiliation{CAS Key Laboratory of FAST, National Astronomical Observatories, CAS, Beijing 100012, China}
\affiliation{NAOC-UKZN Computational Astrophysics Centre (NUCAC), University of KwaZulu-Natal, Durban 4000, South Africa}

\author{Fanyi Meng}
\affiliation{I. Physikalisches Institut, Universit\"{a}t z\"{u} K\"{o}ln, Z\"{u}lpicher Str. 77, D-50937 K\"{o}ln, Germany}

\author{Tianwei Zhang}
\affiliation{I. Physikalisches Institut, Universit\"{a}t z\"{u} K\"{o}ln, Z\"{u}lpicher Str. 77, D-50937 K\"{o}ln, Germany}

\author{Mengyao Tang}
\affiliation{Department of Astronomy, Yunnan University, and Key Laboratory of Astroparticle Physics of Yunnan Province, Kunming 650091, China, chaozhangchz@163.com, slqin@bao.ac.cn}

\author{Lixia Yuan}
\affiliation{National Astronomical Observatories, Chinese Academy of Sciences, 20A Datun Road, Chaoyang District, Beijing 100012, China}

\author{Chenlin Zhou}
\affiliation{National Astronomical Observatories, Chinese Academy of Sciences, 20A Datun Road, Chaoyang District, Beijing 100012, China}

\author{Jarken Esimbek}
\affiliation{Xinjiang Astronomical Observatory, Chinese Academy of Sciences, Urumqi 830011, China}
\affiliation{Key Laboratory of Radio Astronomy, Chinese Academy of Sciences, Urumqi 830011, China}

\author{Yan Zhou}
\affiliation{Department of Astronomy, Yunnan University, and Key Laboratory of Astroparticle Physics of Yunnan Province, Kunming 650091, China, chaozhangchz@163.com, slqin@bao.ac.cn}

\author{Ping Chen}
\affiliation{Department of Astronomy, School of Physics, Peking University, Beijing, 100871, China, ywu@pku.edu.cn}
\affiliation{Kavli Institute for Astronomy and Astrophysics, Peking University, Beijing 100871, China}

\author{Runjie Hu}
\affiliation{Department of Astronomy, School of Physics, Peking University, Beijing, 100871, China, ywu@pku.edu.cn}

\begin{abstract}
Sixty five Planck Galactic cold clumps (PGCCs) from the first quadrant (IQuad) and thirty nine of PGCCs from the Anti-Center direction region (ACent) were observed in $^{12}$CO, $^{13}$CO and C$^{18}$O J=1-0 lines using the PMO 13.7-m telescope. All the targets were detected with all the three lines, except for 12 IQuad and 8 ACent PGCCs without C$^{18}$O detection. Seventy six and 49 velocity components were obtained in IQuad and ACent respectively. One-hundred and forty-six cores were extracted from 76 IQuad clumps and 100 cores from 49 ACent clumps. The average T$_{\mathrm{ex}}$ of IQuad cores and ACent cores are 12.4 K and 12.1 K, respectively. The average line width of $^{13}$CO of IQuad cores and ACent cores are 1.55 km s$^{-1}$ and 1.77 km s$^{-1}$, respectively.
Among the detected cores, 24 in IQuad and 13 in ACent have asymmetric line profiles. The small blue excesses, $\sim$0.03 in IQuad and 0.01 in ACent, indicate that the star formation is not active in these PGCC cores.
Power-law fittings of core mass function to the high mass end give indexes of --0.57 in IQuad and --1.02 in ACent which are flatter than the slope of initial mass function given by \citeauthor{1955ApJ...121..161S}. The large turnover masses with value of 28 M$_{\bigodot}$ for IQuad cores and 77 M$_{\bigodot}$ for ACent cores suggest low star formation efficiencies in PGCCs. The correlation between virial mass and gas mass indicates that most of PGCC cores in both regions are not likely pressure-confined.

\end{abstract}


\section{Introduction} \label{sec:intro}
Examining the initial conditions is essential for studying star formation. Astronomers have been searching for gravitationally bound starless cores prior to protostar phase \citep[also called prestellar cores;][]{1994MNRAS.268..276W,2016MNRAS.463.1008W} and obtaining their parameters \citep{1983ApJ...266..309M}, because it is believed that the property of prestellar cores is the key to answer some important questions about star formation such as the initial mass function \citep{1979ApJS...41..513M}. But progress is still limited.
The first quadrant (hereafter IQuad, 10$^{\circ}\le$ l $\le$ 100$^{\circ}$, --10$^{\circ}\le$ b$\le$ 10$^{\circ}$) and the Anti-Center direction region (hereafter ACent, 175$^{\circ}\le$ l $\le$ 210$^{\circ}$, --9$^{\circ}\le$ b$\le$ 7$^{\circ}$) are two of the most widely studied molecular regions in our Galaxy, which were first observed by \citet{1987ApJ...322..706D,1985PhDT.........3H} respectively. 
Star formation activities such as outflows were frequently reported in molecular complexes of Cygnus X and Monoceros OB1 located in IQuad and ACent respectively \citep[e.g.,][]{1991ApJ...374..540G,1992ApJS...81..267L,2003MNRAS.344..809W,2003AJ....125..842W,2012A&A...541A..79G}.
In IQuad, far infrared (FIR) 160 $\mu$m and 260 $\mu$m emission as well as H110$_{\alpha}$ recombination line observations revealed that FIR sources are coincident with H\,{\footnotesize II}\ regions and associated with molecular clouds \citep{1983BAAS...15..941M}. Triggered star formation has been observed in Vul OB1 in IQuad \citep{2001A&A...374..682E,2010ApJ...712..797B}.
In ACent, \emph{Spitzer} aperture photometry toward Mon OB1 East giant molecular cloud (GMC) shows that the ratio of gas column density to young stellar object (YSO) surface density is large, and thus the undergoing star formation is suggested by \citet{2014ApJ...794..124R}.
However, most of previous works about the two regions focused on particular regions with active star formation. Our understanding of initial conditions of star formation are limited. A study of prestellar cores is thus urgently needed.

The \emph{Planck} satellite all-sky survey provided us Planck Galactic cold clump (PGCC) sample \citep{2011A&A...536A..23P}. PGCCs are cold with dust temperatures ranging from 10 to 15 K and have modest column densities of $N_{\mathrm{H_2}}$ from 0.1 to 1.6 $\times 10^{22}$ cm$^{-2}$.
To investigate the properties of molecular clouds associated PGCCs, a survey towards 674 PGCCs was carried out soon after the releasing of the Early Cold Core Catelogue \citep[ECC;][]{2011A&A...536A...7P} by the PMO 13.7 m telescope in the J=1-0 transitions of $^{12}$CO, $^{13}$CO and C$^{18}$O in single point mode \citep{2012ApJ...756...76W}. 
The well correlated centroid velocities of the three lines and narrow line widths (with $\Delta V_{\mathrm{^{12}CO}}$ typically smaller than 2 km s$^{-1}$) indicate that PGCCs are not very active. However, about ten percent of spectra with asymmetric profiles, indicative of star formation activity, are existent in PGCCs. 
Large sample of observations using dense gas tracers HCN and HCO$^+$ toword PGCCs showed that intensities of HCN and HCO$^+$ are well correlated, while no correlation was found between dense gas tracers and CO \citep{2016ApJ...820...37Y}. 
The observations also showed that some of PGCCs are directly associated with dense cores in prestellar phases of star formation \citep{2011A&A...536A..23P,2016A&A...594A..28P,2015PKAS...30...79L,2016ApJS..222....7L,2018ApJS..234...28L,2018ApJ...859..151L,2018ApJ...856..141T,2019A&A...622A..32L}. 
The characteristics of coldness and enough compactness make them good targets to investigate the initial conditions of star formation \citep{2012ApJ...756...76W,2010A&A...518L..93J,2012A&A...541A..12J,2015A&A...584A..93J,2017ApJS..228...12T}.

In this paper, we report the results of a $^{12}$CO(1-0), $^{13}$CO(1-0) and C$^{18}$O(1-0) survey towards 104 PGCC sources, 65 of which are located in the IQuad and others in the ACent. These PGCCs are selected to investigate the properties of PGCCs as well as their evolutionary stages in the two regions. We describe our samples in section 2.1 and observations in section 2.2. The results are presented in section 3. Detailed discussion is followed in section 4. We summarize main conclusions in section 5.

\section{Samples and Observations} \label{sec:obs}
\subsection{Samples}
\begin{figure*}[htp]
	\centering
	\includegraphics[width=1.2\linewidth,angle=90]{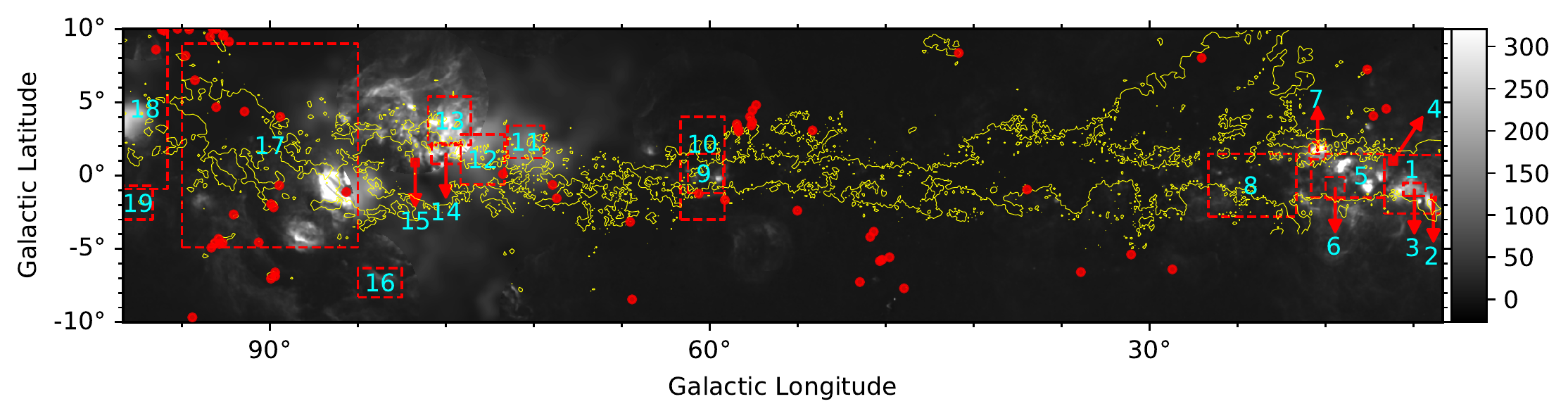} 
	\includegraphics[width=1\linewidth,angle=90]{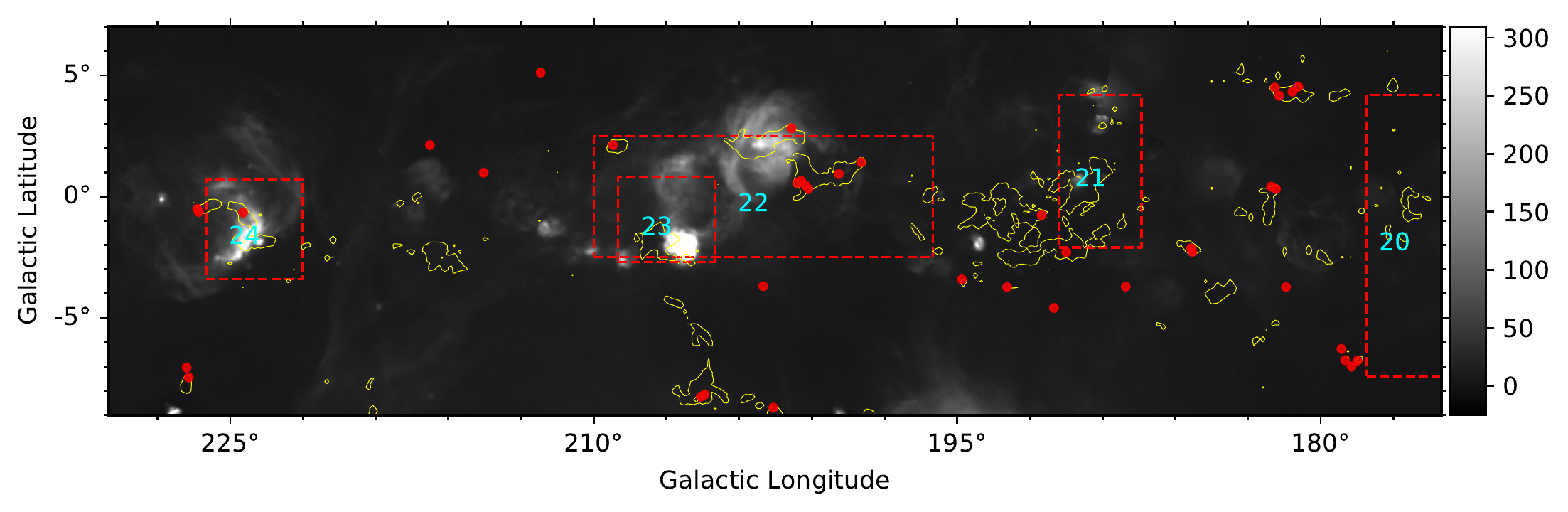} 
	\caption{Longitude-latitude position of 104 PGCCs. The sample sources were marked as red dots. The background gray scale map is the intensity map of H$\alpha$ \citep{2003ApJS..149..405H}. The contour is the velocity integrated CO intensity map in the velocity range of [-50, 50] km s$^{-1}$ from Columbia 1.2 m CO survey \citep{1987ApJ...322..706D}. The contour level is 5$\times$2 K km s$^{-1}$ (1$\sigma$). Twenty-four OB associations are labeled as red dashed rectangles. No. 1-24 OB associations are SGR-OB1, SGR-OB7, SGR-OB4, SGR-OB6, Ser-OB1, SCT-OB3, Ser-OB2, SCT-OB2, Vul-OB1, Vul-OB4, Cyg-OB3, Cyg-OB1, Cyg-OB8, Cyg-OB9, Cyg-OB2, Cyg-OB4, Cyg-OB7, Cep-OB2, Cep-OB1, Aur-OB1, Gem-OB1, Mon-OB1, Mon-OB2 and CMa-OB1, respectively. The limits of the OB associations are from \citet{1978ApJS...38..309H}.}
	\label{Galactic-plane}
\end{figure*}

In order to obtain an overview of the northern sky PGCCs after combining the IIQuad (98$^{\circ}\le$ l $\le$ 180$^{\circ}$, --4$^{\circ}\le$ b $\le$ 10$^{\circ}$) PGCCs \citep{2016ApJS..224...43Z}, we broaden the longitude scope of ACent to 175$^{\circ}$--230$^{\circ}$. IQuad includes some sub-regions, such as Serpens, Aquila, Vulpecula, Cygnus, Cepheus, Cloud A and Cloud B \citep{1987ApJ...322..706D}. Sub-regions of Auriga, Gemini, Monoceros and Canis Major are included in ACent. The sample in this paper are a subset of 674 Planck Galactic cold clumps (PGCCs) observed by \citet{2012ApJ...756...76W} located in IQuad and ACent, selected from the Early Cold Core Catelogue \citep[ECC;][]{2011A&A...536A...7P}. Sixty-five IQuad PGCCs and 39 ACent PGCCs were selected. The names, coordinates and locations of 25 samples are listed in Table \ref{samples}. The longitude-latitude positions of the sample sources are shown in Figure \ref{Galactic-plane}.

\subsection{Observations}
Observations of 65 IQuad PGCCs and 39 ACent PGCCs were carried out with the 13.7 m millimeter-wavelength telescope of Purple Mountain Observatory (PMO) in the J=1-0 transition of $^{12} $CO, $^{13} $CO and C$^{18}$O from April to May 2011 and December 2011 to January 2012. The $3\times3$-beam sideband separation Superconduction Spectroscopic Array Receiver system was used as the front end \citep{2012ITTST...2..593S}. 
The HPBW is 52 arcsec in the 115 GHz band.
The mean beam efficiency is about $50\%$. The pointing accuracy is better than $5''$. The fast Fourier transform spectrometers (FFTSs) were used as the backend. Each FFTS with a total bandwidth of 1 GHz provides 16384 channels. The velocity resolutions are 0.16 km s$^{-1}$ for $^{12}$CO, and 0.17 km  s$^{-1}$ for $^{13}$CO and C$^{18}$O. The $^{12}$CO emission line was observed in the upper sideband (USB) with the system temperature (T$_{sys}$) around 210 K, while the $^{13}$CO and C$^{18}$O emission lines were observed simultaneously in the lower sideband (LSB) with T$_{sys}$ around 120 K. The typical rms was 0.2 K in T$_A^*$ for $^{12}$CO, and 0.1 K for $^{13}$CO and C$^{18}$O. The on-the-fly (OTF) observing mode was applied. The antenna continuously scanned a region of $22'\times 22'$ centroid on the PGCCs with a scanning rate of $20''$ per second. Because of the high rms noise level at the edges of OTF maps, only the central $14'\times 14'$ regions were used for analyses in this work. The data were reduced by CLASS and GREG in the GILDAS package \citep{2000ASPC..217..299G, 2005sf2a.conf..721P}. All statistical figures were plotted by open source Python package.

\section{Results} \label{sec:results}
\begin{figure*}
	\includegraphics[width=0.7\linewidth]{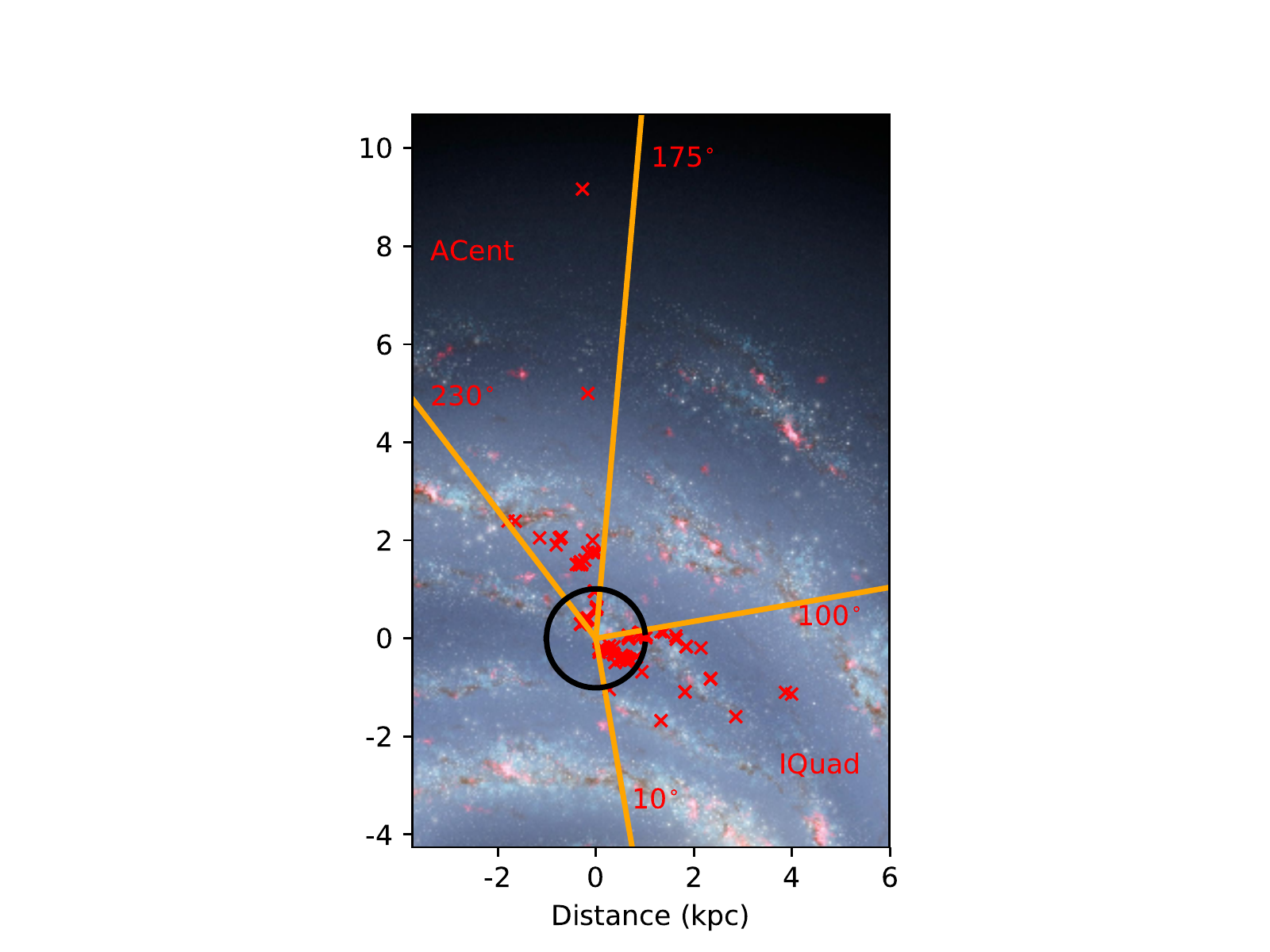}
	\caption{Galactic face-on position of 125 clumps. The clumps were marked as red crosses. The Solar system was defined as the origin of coordinate. The x and y axes shows the distance in Kpc. The black circle is plotted as the 1 Kpc distance from the Sun. Orange lines are the Galactic longitude.}
	\label{face-on}
\end{figure*}

CO (1-0) and $^{13}$CO (1-0) emission lines were detected in all 65 IQuad PGCCs and 39 ACent PGCCs. C$^{18}$O emssions were detected in the 53 ($\sim$ 81.5\%) IQuad PGCCs and in the 31 ($\sim$ 79.5\%) IQuad PGCCs.
The detection rates of C$^{18}$O in both regions are comparable with those in Taurus \citep[77.5\%;][]{2013ApJS..209...37M} and IIQuad \citep[84.4\%;][]{2016ApJS..224...43Z}.
We noted that some PGCCs have two or multiple velocity components. In the case where two peaks are blended, we checked the integrated maps of the blue and red components. If the distributions of the two integrated maps are not highly coherent, we confirm that the two peak components were two velocity components.
Eight IQuad PGCCs have double velocity components and one IQuad PGCC has quadruple velocity components. Ten ACent PGCCs have double velocity components. In total, 76 and 49 velocity components were obtained in IQuad and ACent, respectively. The integrated intensity maps of different velocity components show dissimilar features. We thus considered each velocity component as a separate clump. Centroid velocities of the clumps were obtained by Gaussian fitting. On the basis of the Milky way rotation curve, the spiral arm model and the giant molecular clouds with a measured parallax, if the (l, b, v) coodinates of the sources are given, the distance distributions in light-of-sight direction of the sources can be obtained by \emph{Bayesian distance calculator}\footnote{http://bessel.vlbi-astrometry.org/bayesian} \citep{2016ApJ...823...77R}.
The kinematic distances were derived with the highest probabilities of the distance distributions.
Galactic face-on positions of the 125 clumps are shown in Figure \ref{face-on}.

\begin{figure*}
	\includegraphics[width=155mm,height=200mm]{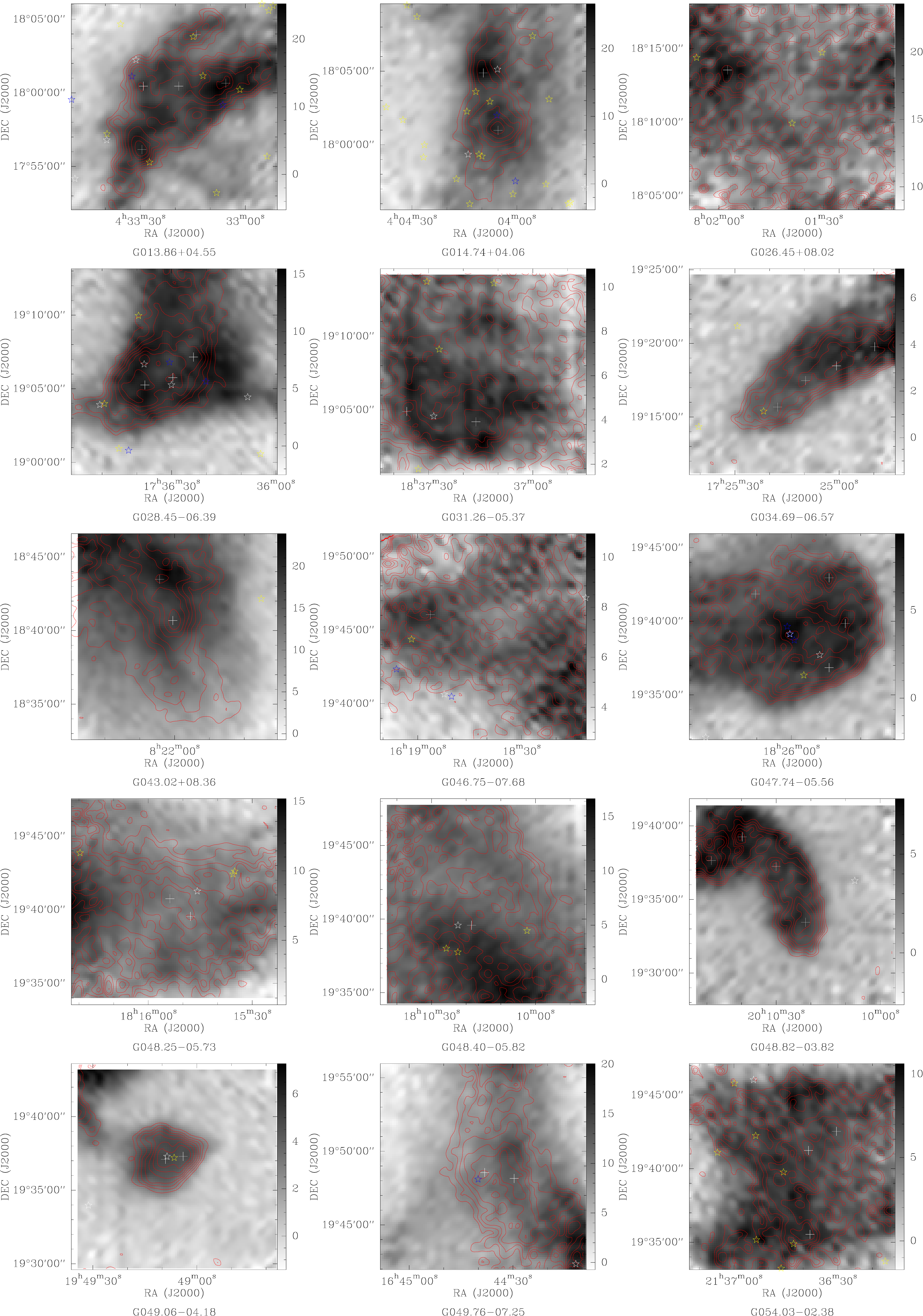}
	\caption{The velocity integrated intensity maps. The integrated intensities of $^{13}$CO are represented as red contours stepped from 30\% to 90\% by 10\% of the peak value. Gray scale is the integrated intensity of $^{12}$CO. The identified core positions are marked as white "+". The blue, yellow and white stars presented the Class I/II YSO candidates, Class III YSO candidates and \emph{IRAS} pointing sources respectively. The Class I/II and III YSO candidates are selected from the \emph{WISE All-Sky survey} catalogues presented by \citet{2016MNRAS.458.3479M}. The \emph{IRAS} pointing source catalogue are presented by \citet{1988SSSC..C......0H}.
	\label{maps}
	}
	\tablecomments{An extended, color version of this figure is available in the online journal.}
\end{figure*}

The integrated intensity maps of CO are plotted and 15 of them are presented as sample maps in Figure \ref{maps}, in which the clumps show different morphologies,  such as isolated features, filamentary structures and chains of dense cores. Eight clumps of G015.16+07.23a, G034.69-06.57, G048.82-03.82, G056.84+04.81, G093.22+04.59, G188.04-03.71, G191.00-04.59 and G198.96+01.41b show filamentary structures with sausage shape. 
In this paper, we mainly focus on the dense cores, which have an isoline of 50\% $^{13}$CO peak intensity and are denser than the surrounding regions in the same clump. These cores are starless cores if there is no embedded stars \citep{1989ApJS...71...89B}. If the cores are bound by gravity, we refer to prestellar cores.
Components of a PGCC are denoted as ``a, b, c, ...", while the cores are labeled as ``1, 2, 3, ...", following \citet{2012ApJ...756...76W}. The velocities of the different components (V$_a$, V$_b$, V$_c$, ...) in one PGCC must have V$_a<$V$_b<$ V$_c$.
Thirty-six clumps contain one dense core and 73 contain more than one dense cores. In 14 maps, CO emission is too diffuse to identify dense core.
In total, 146 cores were identified in 70 IQuad clumps, and 100 cores in 39 ACent clumps.

\subsection{Line Parameters} \label{subsec:obspara}
Gaussian fittings to the three CO lines for every velocity component at each core centers were made. The centroid velocity ($V_{lsr}$),
main beam brightness temperature (T$_b$), and FWHM ($\Delta$V) were obtained and listed in Table \ref{obs-para}. 
The centroid velocity of $^{13}$CO  J=1-0 line of each core was adopted as the systemic velocity, because 
$^{12}$CO J=1-0 line is optically thick and C$^{18}$O J=1-0 line has low signal to noise ratio. 

The statistical properties of $V_{lsr}$, $\Delta$V and T$_b$ of the three CO lines, including the ranges of values, the mean values, the medians and the standard deviations, are listed in Table \ref{stat-obs-para}. 
Both the median absolute deviations (MADs) of line width for IQuad cores and ACent cores are 0.42 km/s in $^{13}$CO lines, which is large enough to affect the comparison of the median value between the two regions, suggesting that there are no obvious differences in line width between the two regions.

\subsection{Line Profiles} \label{subsec:profiles}
\begin{figure*}[htp]
	\includegraphics[width=170mm]{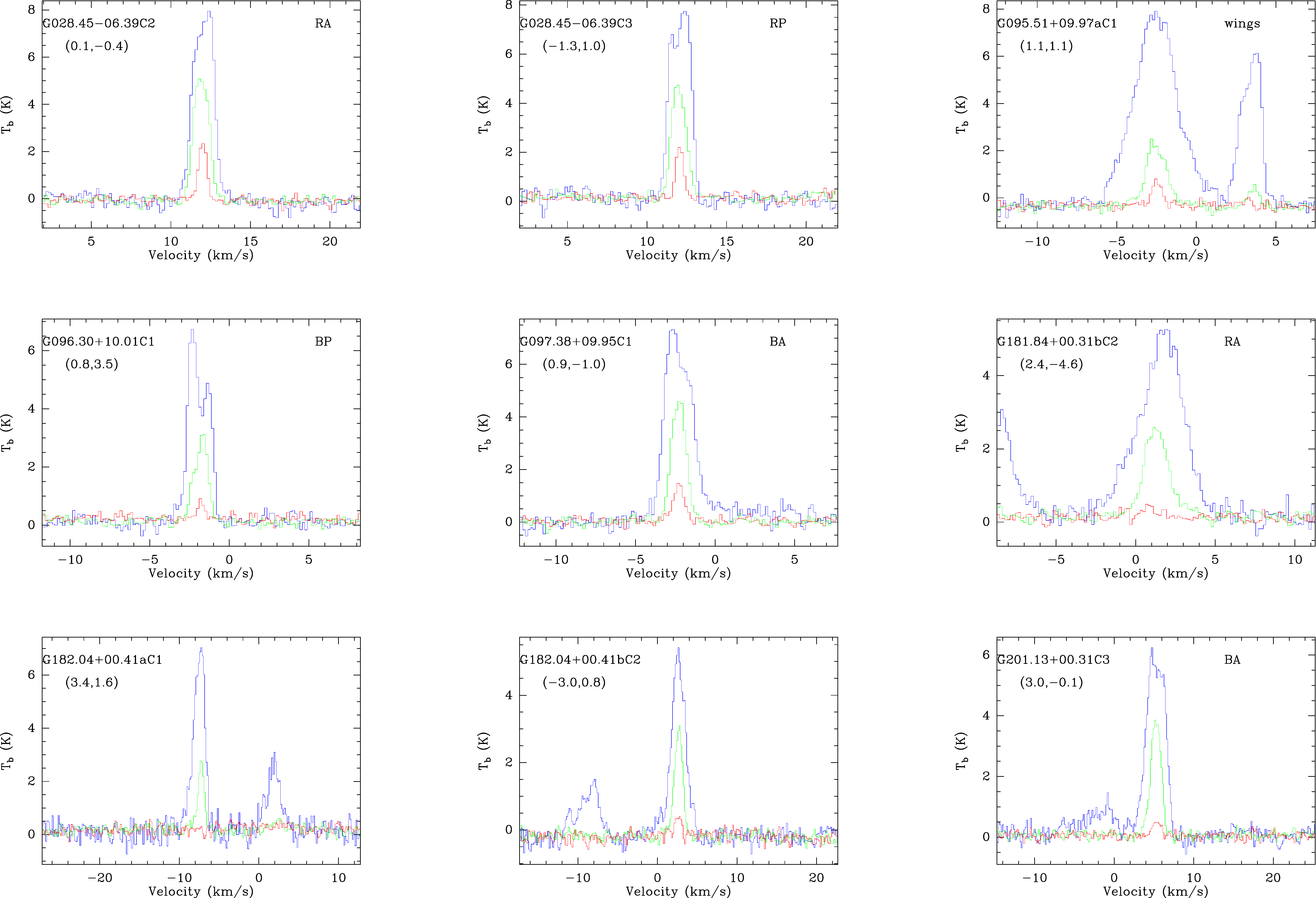}
	\caption{Sample spectra extracted from peak position of the cores. For each plot, the three lines of $^{12}$CO, $^{13}$CO and C$^{18}$O are colored blue, green and red, respectively. The core name and the offset (in arcmin) between the core center and clump center list in the top-left corner of each plot. The asymmetric line profiles of blue profile (BP), red profile (RP), blue asymmetry (BA), red asymmetry (RA) and wings, if any, list in the top-right corner of each plot.
	\label{spec}
	}
	\tablecomments{An extended, color version of this figure is available in the online journal.}
\end{figure*}

The CO, $^{13}$CO and C$^{18}$O spectra are extracted from the peak positions of the dense cores. Figure \ref{spec} presents sample spectra of 9 cores. 
One can see that the spectra of G028.45-06.39C3 and G096.30+10.01C1 show double-peaked $^{12}$CO emission lines and single-peaked $^{13}$CO and C$^{18}$O emission lines, and the $^{13}$CO and C$^{18}$O lines peak at absorption dips of the $^{12}$CO lines. The line shoulders are found in the $^{12}$CO lines toward G028.45-06.39C2, G097.38+09.95C1, G181.84+00.31bC2 and G201.13+00.31C3. The spectra of G095.51+09.97aC1, G181.84+00.31bC2, G182.04+00.41aC1 and G182.04+00.41bC2 have two velocity components. Wide $^{12}$CO line wings are found toward the left velocity component of G095.51+09.97aC1.

For the cores with double-peaked optically thick lines, if the optically thin lines peak at absorption dip of the optically thick lines, we can identify the double-peaked optically thick lines as red- or blue-asymmetric profile in case that the blue peaks are higher or lower than the red peaks respectively. The blue or red profile indicates inward or outward motion \citep{1992ApJ...394..204Z}. Usually the dense gas tracers such as HCO$^+$, HCN, CS and N$_2$H$^+$ are used for identifying collapse in the compact cores \citep{1997ApJ...489..719M,2005ApJ...620..800D,2007prpl.conf...17D,2010MNRAS.407.2434S}. Observations suggested that PGCCs have low temperature and modest density \citep{2011A&A...536A..23P,2016A&A...594A..28P}, and then dense gas tracers of HCO$^+$, HCN in PGCCs are difficult to excite and have lower detection rate than $^{12}$CO and $^{13}$CO \citep{2016ApJ...820...37Y}. In the past, intermediate-density gas tracers of $^{12}$CO and $^{13}$CO were employed to observe the infall motion in both low mass and high mass star forming regions \citep[e.g.,][]{1975ApJ...199...79K,1977ApJ...211..122S,2006ApJ...640..842R,2008MNRAS.391..869G,2009ApJ...697L.116W,2010ApJ...710..843S}. Therefore, we can identify blue and red line profiles by use of the optically thick $^{12}$CO line and optically thin $^{13}$CO line (see Sect. \ref{derived-para}). The blue profile can be verified by asymmetric parameter $\delta$V$<-$0.25 in which $\delta$V = (V$_{thick}$-V$_{thin}$)/$\Delta$V$_{thin}$ \citep{1997ApJ...489..719M}, where V$_{\rm thick}$ is the velocity of the $^{12}$CO line peak (V$_{\rm peak}$($^{12}$CO)), V$_{\rm thin}$ is the centroid velocity of the $^{13}$CO line (V$_{\rm lsr}$($^{13}$CO)), and
$\Delta$V$_{\rm thin}$ is the line width (FWHM) of the $^{13}$CO line. The red profile would have $\delta$V$>0.25$. In total, 8 blue profiles and 3 red profiles were identified in 146 IQuad cores, and 2 blue profiles and 1 red profiles were identified in 100 ACent cores.

Some $^{12}$CO line peaks are merely skewed and show line shoulders. The $\delta$V$_{12}$ is defined as (V$_{thick}$-V$_{lsr}(^{12}\mathrm{CO})$)/$\Delta$V$_{thin}$ \citep{2012ApJ...756...76W} to identify these skewed lines as blue or red asymmetries if $\delta$V$_{12}$ is smaller than --0.25 or larger than 0.25 respectively. Five blue asymmetries and 6 red asymmetries were identified in IQuad, and 5 blue asymmetries and 5 red asymmetries were identified in ACent. In total, there are 35 asymmetric profiles detected. 

\begin{figure*}
    \centering
	\includegraphics[width=0.45\linewidth,height=0.3\linewidth]{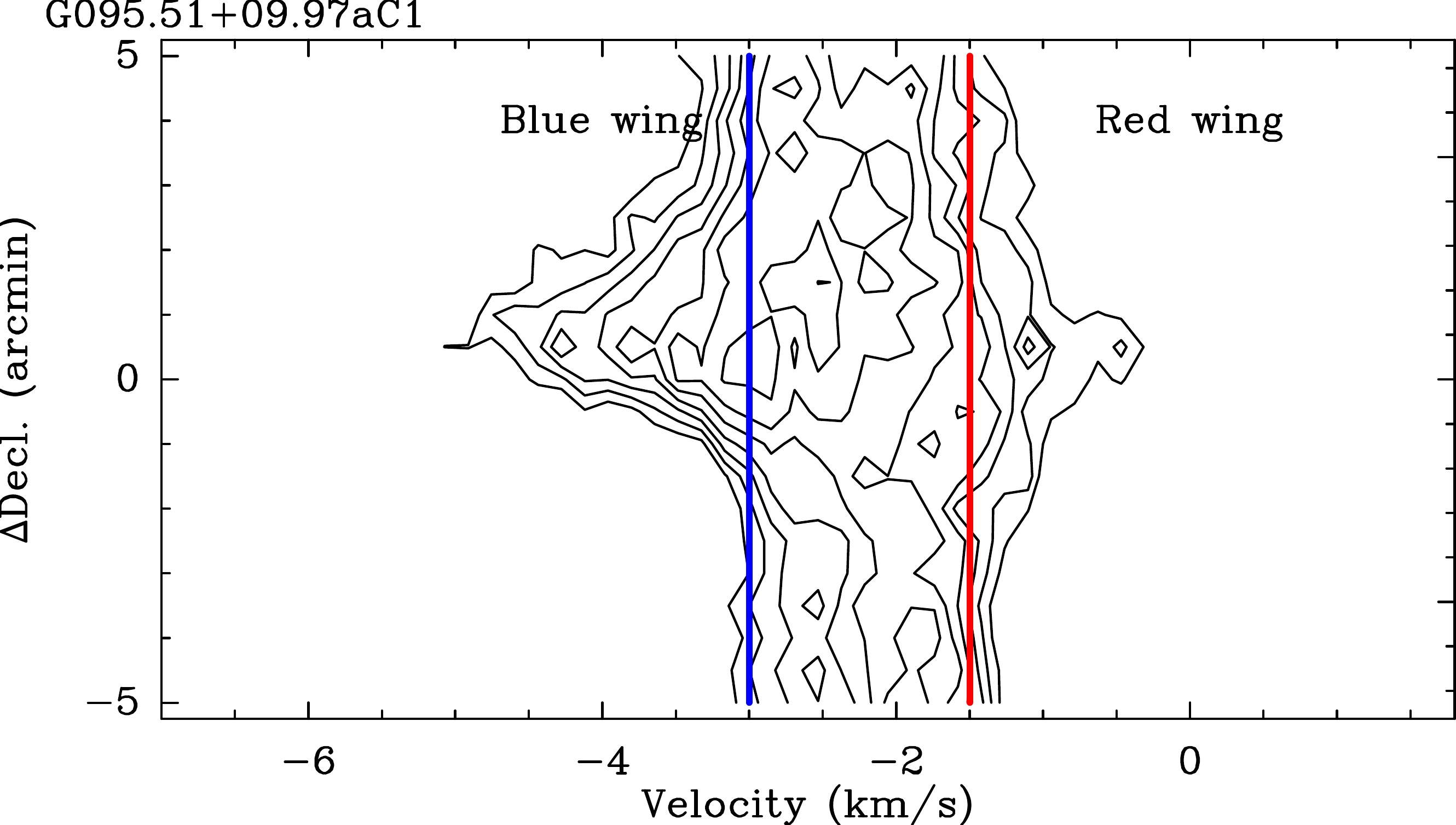}
	\includegraphics[width=0.45\linewidth,height=0.3\linewidth]{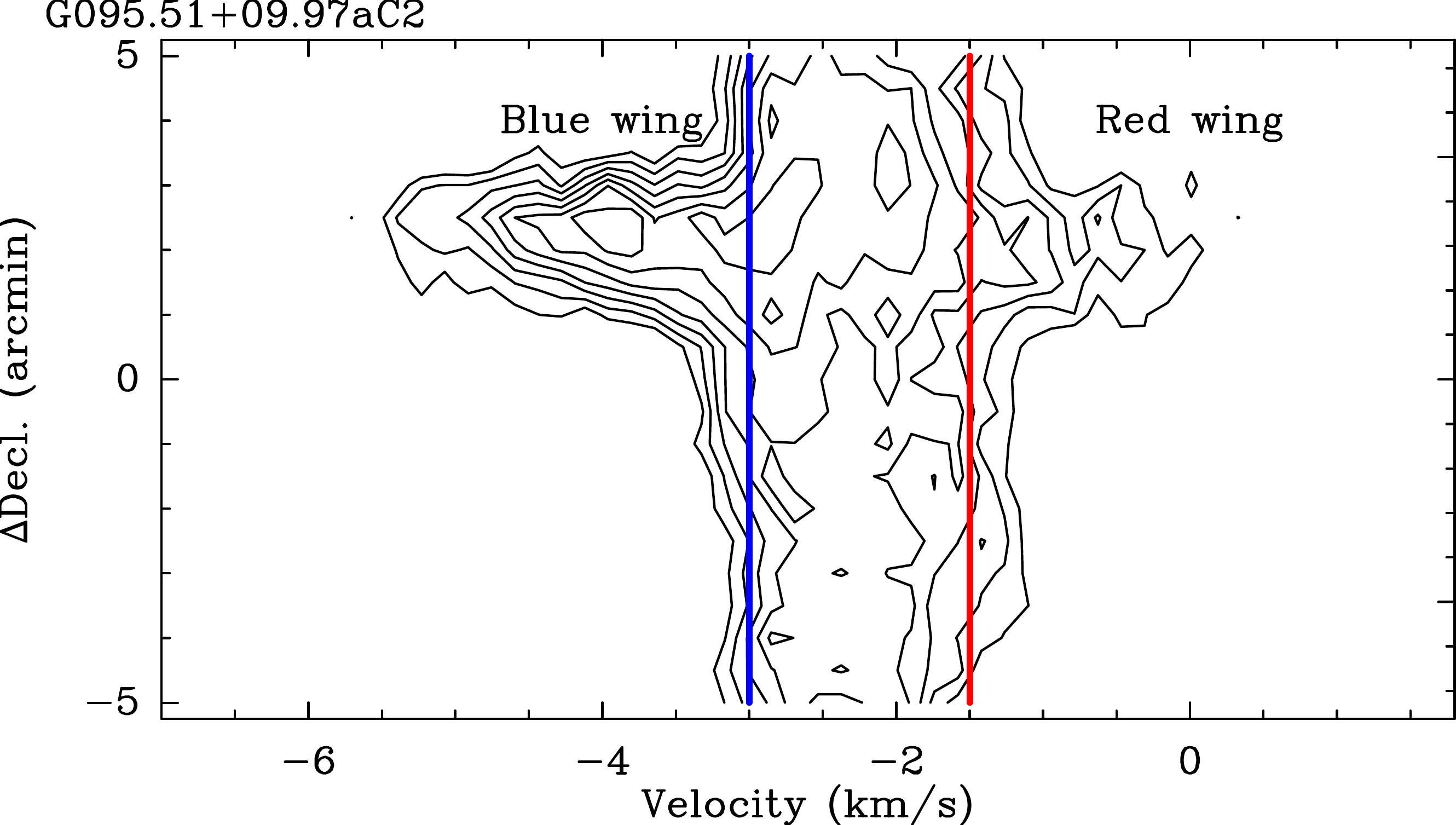}
	\caption{$P-V$ diagram of the two cores. Left is taken from R.A. offset 1.5$'$, decl. offset from --5$'$ to 5$'$ in G095.51+09.97aC1; Right is taken from R.A. offset --1$'$, decl. offset from --5$'$ to 5$'$ in G095.51+09.97aC2. The possible outflow motions are labeled.
	\label{P-V}
}
\end{figure*}

Of the detected cores in both regions, 8 were found having wing(s). The P-V diagrams of the two cores (G095.51+09.97aC1 and G095.51+09.97aC2 which resolved from the same clump) show convex isolines (see Figure \ref{P-V}). The core components have constant velocity around systemic velocity, while high-velocity wings show convex isolines and velocity gradient with respect to systemic velocity in the P-V diagram. The high-velocity wings may origin from the outflows \citep{1985ARA&A..23..267L,2005AJ....129..330W}. Therefore we identified two cores as the outflow candidates. 
The spectral profile classfications are listed in last column of Table \ref{obs-para}.

\subsection{Derived Parameters} \label{subsec:derpara}
The excitation temperatures (T$_{\mathrm{ex}}$) and optical depths ($\tau$) can be derived from the solution of the radiation transfer equation
\begin{eqnarray}
&T_b =& \frac{h\nu}{k}\left[\frac{1}{exp(h\nu/kT_{ex})-1} - \frac{1}{exp(h\nu/kT_{bg})-1}\right] \nonumber\\
&&\times [1-exp(-\tau)]f,
\end{eqnarray}
where T$_{bg}$ is the backgroud temperature (2.73 K). Under assumptions that $^{12}$CO J=1-0 is optically thick ($\tau \gg$ 1), beam filling factor $f$ equals to 1, the excitation temperature can be expressed as
\begin{eqnarray}
&T_{ex} =& \frac{h\nu}{k} ln^{-1}\left\{\left[\frac{kT_b(^{12}CO)}{h\nu}+\frac{1}{exp(h\nu/kT_{bg})-1}\right]^{-1}+1\right\}.
\end{eqnarray}
Under conditions of local thermodynamic equilibrium (LTE), $^{12}$CO and $^{13}$CO have the same excitation temperature. We can obtain optical depth of $^{13}$CO ($\tau_{^{13}\mathrm{CO}}$) by
\begin{eqnarray}
	&\tau_{^{13}\mathrm{CO}} =& -ln\left[1-\frac{T_b(^{13}CO)}{T_b(^{12}CO)}\right].
\end{eqnarray}

The T$_{\mathrm{ex}}$ of the 246 cores range from 7.2 to 23 K with a median value of $\sim$11.5 K with MAD of $\sim$2 K. Only 6 cores (2.5\%) have T$_{ex}>20$ K. 
The dust temperature (T$_{\mathrm{dust}}$) of all PGCCs ranges from 5.8 to 20 K with a median value between 13 and 14.5 K \citep{2016A&A...594A..28P}. 
The similar temperature ranges between T$_{\mathrm{ex}}$ and T$_{\mathrm{dust}}$ indicate that the gas and dust are well coupled.
However, the different median temperature (T$_{\mathrm{dust,median}}>$ T$_{\mathrm{ex,median}}$) imply that the gas might be heated by the dust \citep{1974ApJ...189..441G,1989ApJ...340..307W}. The derived optical depths of $^{13}$CO lines range from 0.1 to 3.8 with an average value of 0.7, indicating that $^{13}$CO lines in most of the PGCC cores are optically thin. Based on abundance gradient of $[^{12}C]/[^{13}C] = (7.5\pm1.9)D_{GC}+7.6\pm12.9$ \citep{1994ARA&A..32..191W}, where D$_{GC}$ is the distance of the source departed from the Galactic Center, the [$^{12}$C]/[$^{13}$C] of the cores in IQuad and ACent is derived and rangs from 59 to 124 with an average value of 75. If we take [$^{12}$CO]/[$^{13}$CO]=59, and the lowest $^{13}$CO optical depth of $\sim$0.1, the $^{12}$CO optical depth is estimated to be 5.9, indicating that $^{12}$CO lines of the cores are optically thick.

The column density of $^{13}$CO (N$_{^{13}\mathrm{CO}}$) and C$^{18}$O (N$_{C^{18}O}$) can be derived by \citep{1978afcp.book.....L,1991ApJ...374..540G}
\begin{eqnarray}
N =&& \frac{3k}{8\pi^3B\mu_d^2} \frac{exp\left({h B J(J + 1)}/{kT_{ex}}\right)}{J + 1}  \nonumber \\
&&\times \frac{T_{ex} + h B/3k}{1 - exp(-h\nu/kT_{ex})} \int\tau_{\nu}dV,
\end{eqnarray}
where B, $\mu_d$ and J are the rotational constant, permanent dipole moment of the molecule and the rotational quantum number of the lower state in the observed transition. 
Adopting the typical abundance ratios, $[{\rm H_2}]/[{\rm ^{13}CO}] = 89\times10^4$
\citep{1980ApJ...237....9M} and $[{\rm H_2}]/[{\rm C^{18}O}] = 7\times10^6$
\citep{1982ApJ...262..590F} for the solar neighborhood, the column density of hydrogen (N$_{\mathrm{H}_2}$) can be calculated. We find that the H$_2$ column density values derived from $^{13}$CO and C$^{18}$O are quiet close to each other, suggesting that both $^{13}$CO and C$^{18}$O are optical thin and the optical depth effect in our calculation can be ignored \citep{2013ApJS..209...37M}. 
Therefore we use $^{13}$CO to calculate H$_2$ column densities, as listed in Table \ref{deri-para}.
The median values of H$_2$ column density of the cores in the two regions are quiet similar, 7.2$\times$10$^{21}$ cm$^{-1}$ for IQuad cores and 7.5$\times$10$^{21}$ cm$^{-1}$ for ACent cores.

The one-dimensional velocity dispersion of $^{13}$CO is given by
\begin{equation}\sigma_{^{13}CO} = \frac{\rm \Delta V_{^{13}CO}}{\sqrt{8\ln 2}}.\end{equation}
The thermal ($\sigma_{Th}$) and non-thermal ($\sigma_{NT}$) velocity dispersions can be calculated by
\begin{equation}\sigma_{Th} = \sqrt{\frac{k T_{ex}}{m_H\mu}},\end{equation}
\begin{equation}\sigma_{NT} = \left[\sigma_{^{13}CO}^2 - \frac{k T_{ex}}{m_{^{13}CO}}\right]^{1/2},\end{equation}
where m$_{\rm H}$ is the mass of atomic hydrogen, $\mu$ is the mean weight of molecule which equals to 2.33 \citep{2008A&A...487..993K}, m$_{^{13}\mathrm{CO}}$ is the mass of molecular $^{13}$CO and $k$ is the Boltzman constant. Then the three-dimensional velocity dispersion $\sigma_{3\mathrm{D}}$ can be calculated by
\begin{equation}\sigma_{3D} = \sqrt{3(\sigma_{Th}^2 + \sigma_{NT}^2)}.\end{equation}

All the above derived parameters are listed in Table \ref{deri-para} and their statistical parameters are listed in Table \ref{stat-deri-para}.  

\subsection{Emission region parameters} \label{subsec:mappara}

The offset peak positions relative to the PGCCs, Gaussian semi-major axis (a) and semi-minor axis (b) of the cores were estimated from the elliptic Gaussian fitting of the isoline with 50\% of the peak integrated intensity. The core radius are calculated as $R =\frac{1}{2}\sqrt{\mathrm{ab}}\ \mathrm{D}$, where D is the distance. Volume density can be derived through n$_{\mathrm{H}_2}$ = N$_{\mathrm{H}_2}$/2R.
We estimated gas mass from
\begin{equation} M = \frac{2}{3}\pi R^2m_{H}\mu N_{H_2}.\end{equation}

Assuming that the cores are gravitationally bound isothermal spheres and supported solely by random motions, we can calculate the virial mass M$_{\mathrm{vir}}$ as \citep{1994ApJ...428..693W}
\begin{equation}M_{vir} = \frac{5R\ \sigma^2_{3D}}{3\gamma \ G}, \end{equation}
where G is the gravitational constant. Assuming the density profile follows $\rho \varpropto$ R$^{-2}$, $\gamma$ equals to 5/3.

In molecular clouds, thermal pressure, turbulence and magnetic field can support the gas against gravitational collapse. The Jeans mass is defined to describe the stability of the molecular core. When gas mass is larger than its Jeans mass, gravitational collapse will happen.
Taking thermal pressure and turbulence into account, the Jeans mass can be expressed as following \citep{2008ApJ...684..395H}
\begin{equation}\frac{M_{Jeans}}{M_{\bigodot}} \approx 1.0 a_J \left(\frac{T_{eff}}{10K}\right)^{3/2}\left(\frac{\mu}{2.33}\right)^{-1/2}\left(\frac{n}{10^4cm^{-3}}\right)^{-1/2},\end{equation}
where a$_\mathrm{J} = 1$ is a dimensionless parameter of the order unity, n is the volume density of H$_2$, and the effective kinematic temperature T$_{\mathrm{eff}}$ is adopted as $\frac{ \mathrm{C_{s,eff}}^2 \mu \mathrm{m_H}}{\mathrm{k}}$. The effective sound speed C$_{\mathrm{s,eff}}$ is adopted as $[(\sigma_{\mathrm{NT}})^2+(\sigma_{\mathrm{Th}})^2]^{1/2}$ to account for support of turbulence.

The radius, volumn density, gas mass, virial mass and Jeans mass are listed in Table \ref{area-para} and their statistics are listed in Table \ref{stat-area-para}. The median values of core size (R) are 0.32 pc and 0.43 pc in IQuad and ACent respectively. The difference of median core size between IQuad and ACent can be ignored since MADs of R have the value 0.14 pc in IQuad and 0.21 pc in ACent. We derived the typical R of 207 cores with D $<$ 2 kpc to be 0.1$-$0.6 pc. The mean masses of IQuad cores (291 M$_{\bigodot}$) and ACent cores (151 M$_{\bigodot}$) are extrmely different from their median values (31 M$_{\bigodot}$ in IQuad and 43 M$_{\bigodot}$ in ACent). Most of the cores with extremely high mass tend to have distances larger than 2 kpc. According to Equation (9), (10) and (11), M, M$_{vir}$ and M$_{Jeans}$ are related to R and thus depend on distance (D). Some distant cores have high mass and can enlarge the mean value but have little effect on the median.

\section{Discussion}
\subsection{Galactic distribution} \label{subsec:distribution}
As shown in Figure \ref{Galactic-plane}, the 104 PGCCs are mostly located on the edge of the CO (1-0) emission detected by \citeauthor{1987ApJ...322..706D} using the Columbia 1.2 m telescope with spatial resolution 8$'$. Most of PGCCs have deviated from the peak positions of the H$\alpha$ emission \citep{2003ApJS..149..405H}. 
Some PGCCs are out of the edge of the CO (1-0) emission region, indicating that additional gas was detected by more sensitive observations using the PMO 13.7 m telescope under the guidance of PGCCs detected by $Planck$ Satellite.
PGCCs G84.79-01.11, G201.84+02.81 and G224.47-00.65 have strong H$\alpha$ emission, indicating that they are associated with the H\,{\footnotesize II}\ region and thus could be more active than others. 

The positions of 125 clumps in the Galactic face-on map are shown in Figure \ref{face-on}, in which the black circle denotes an 1 kpc circle from the Solar system. One can see most of IQuad clumps fall into the circle, but considerable ACent clumps are out of the circle, i.e. IQuad clumps are closer to us. It seems deserving that median values of core mass (M) and core size (R) in IQuad are smaller than those in ACent (see Table \ref{stat-area-para}). The clumps in IQuad and ACent are located at a distance of 0.23--4.15 kpc with a median value of 0.8 kpc and 0.4--9.17 kpc with a median value of 1.5 kpc from the Sun, respectively. Some clumps are situated in the inter-arm region.

\subsection{Emission lines} \label{subsec:el}

\begin{figure*}
    \centering
    \includegraphics[width=0.4\linewidth]{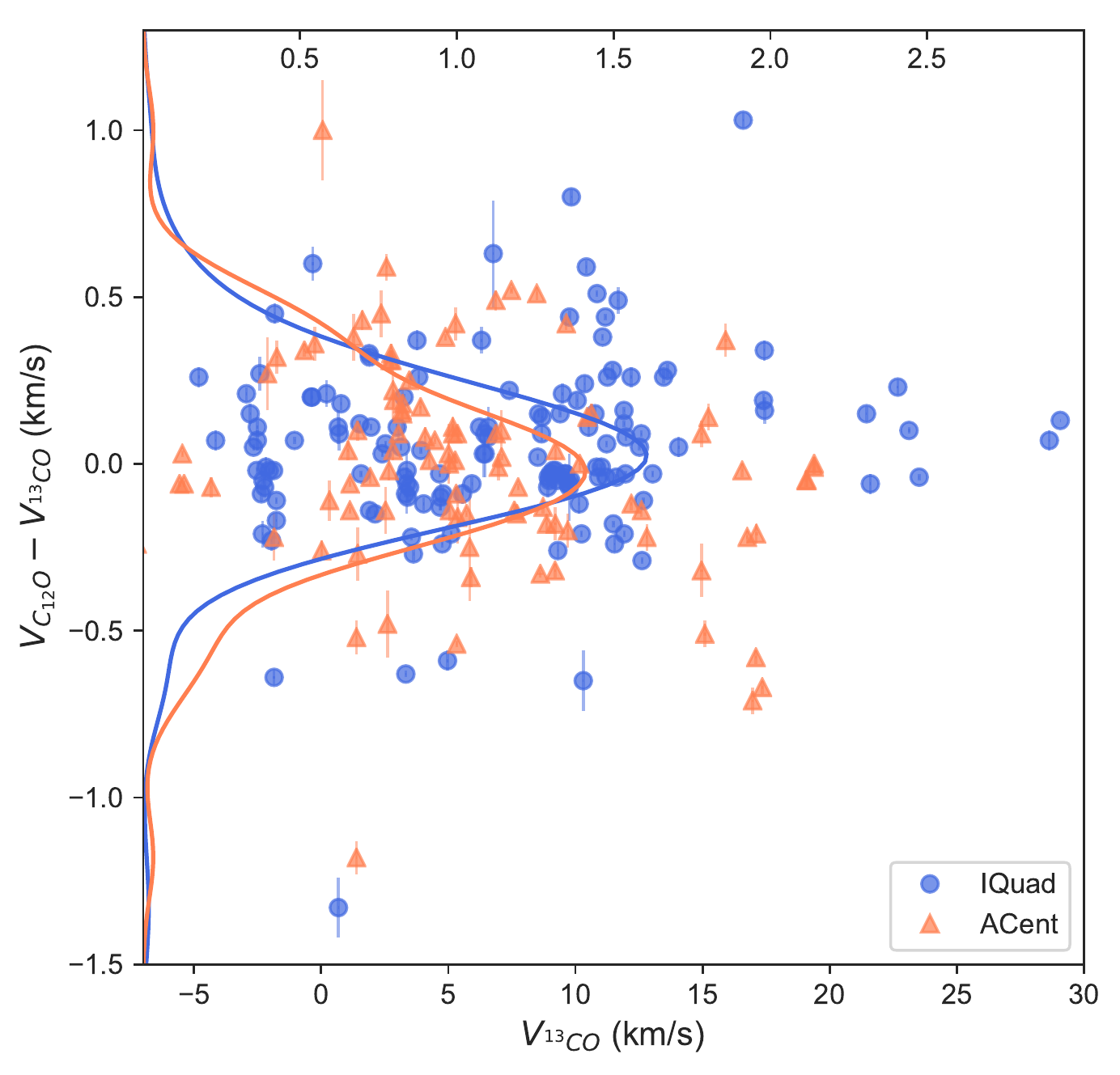}
    \includegraphics[width=0.4\linewidth]{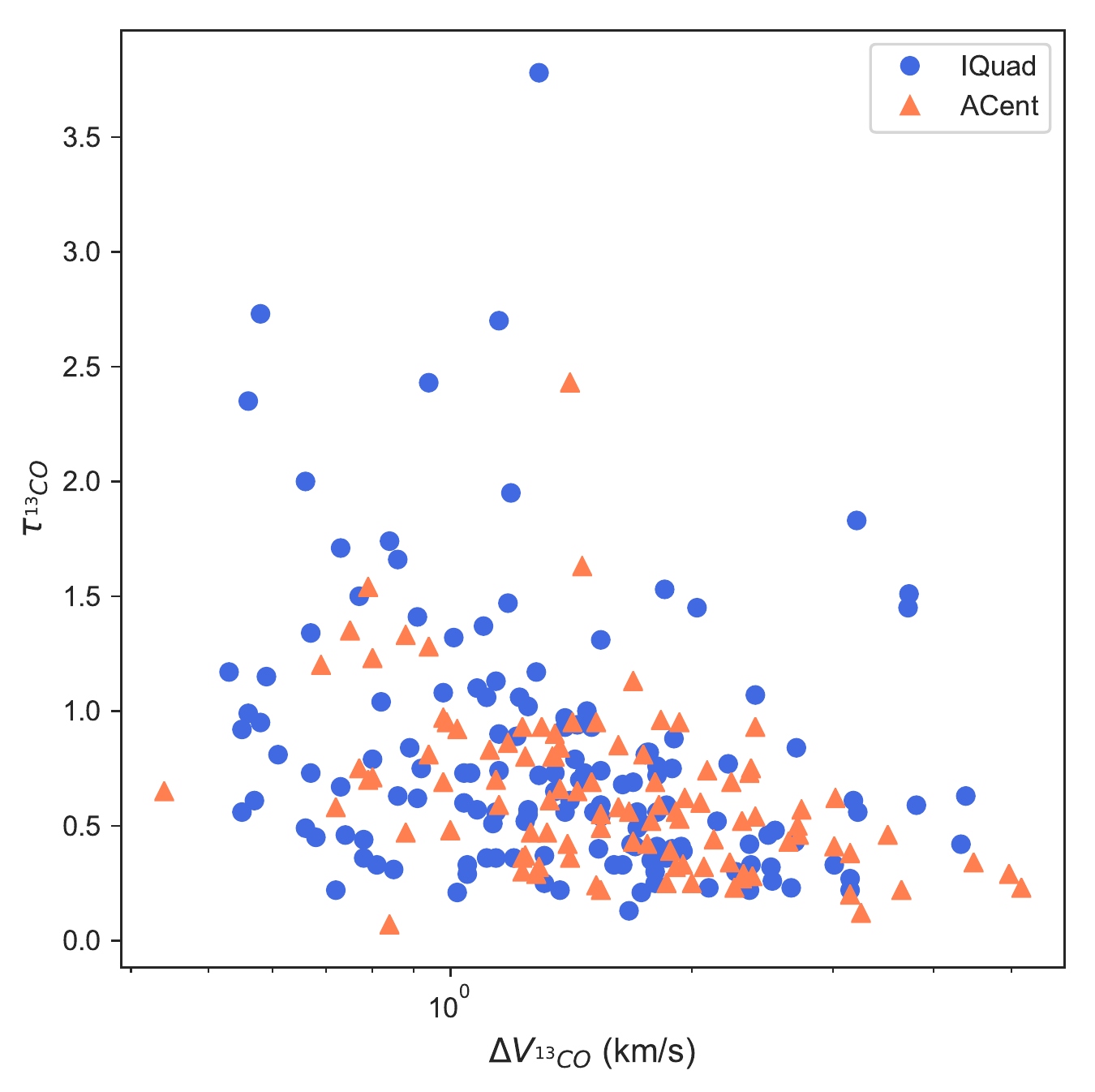}
    \includegraphics[width=0.4\linewidth]{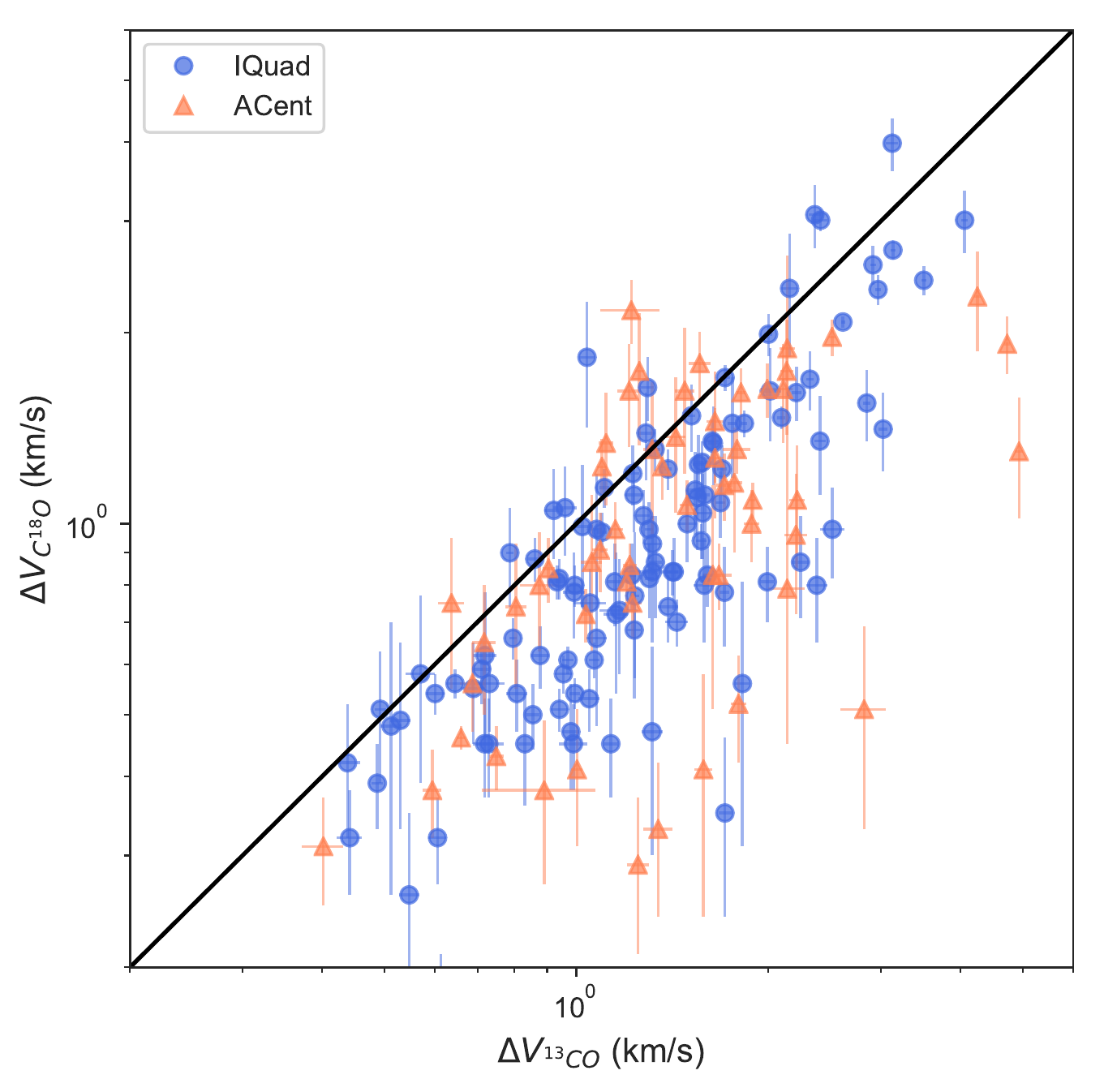}
    \includegraphics[width=0.4\linewidth]{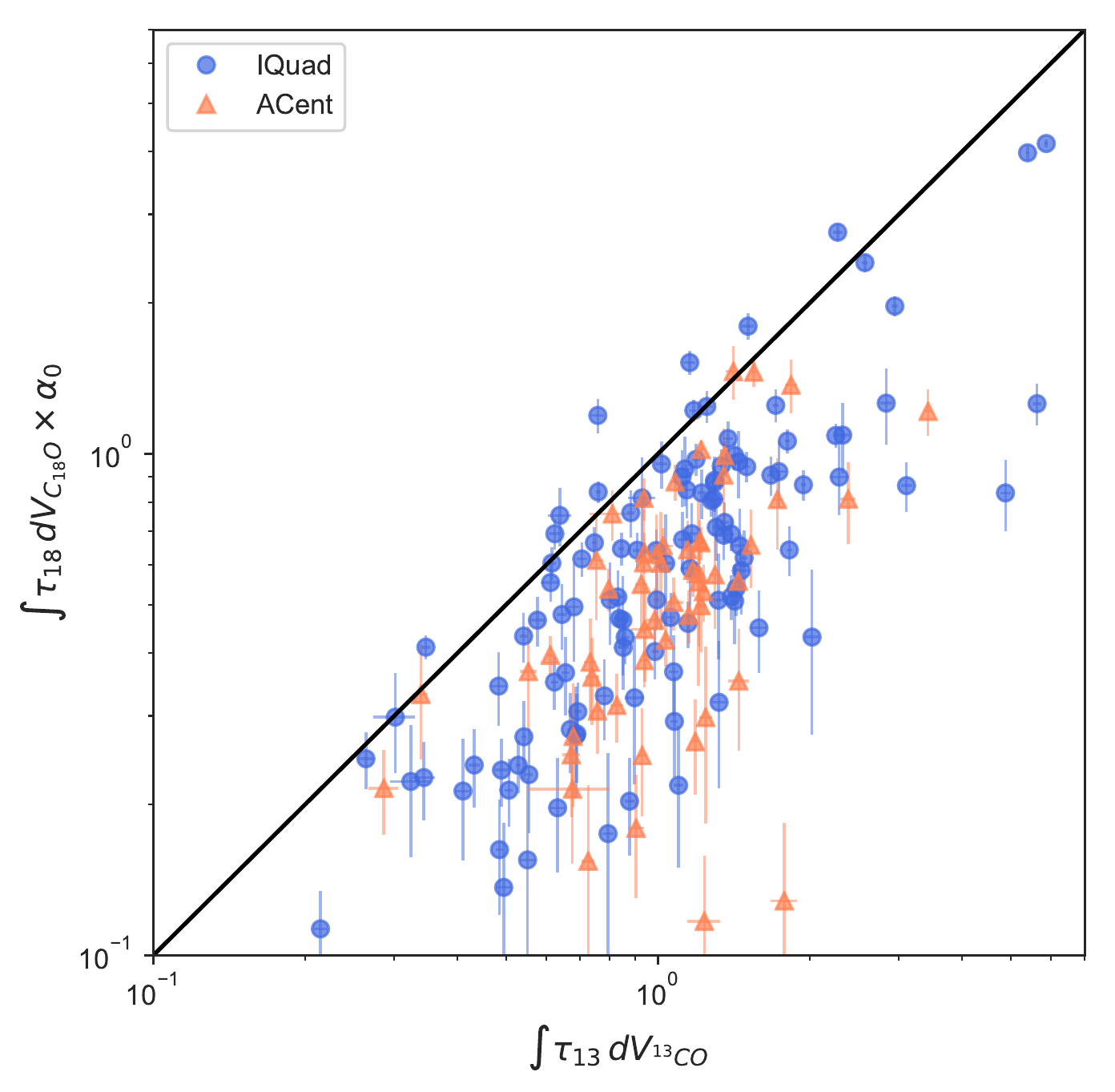}
	\caption{Top left: the centroid velocity of $^{13}$CO emission line (V$_{^{13}\mathrm{CO}}$) v.s. the centroid velocity of C$^{18}$O minus that of $^{13}$CO (V$_{\mathrm{C}^{18}\mathrm{O}}$-V$_{^{13}\mathrm{CO}}$). The curves are the distribution and their colors are as same as the region data symbols. 
	Top right: the line widths of $^{13}$CO emission line ($\Delta V_{^{13}\mathrm{CO}}$) v.s. the line widths of C$^{18}$O emission line ($\Delta V_{\mathrm{C}^{18}\mathrm{O}}$). 
	Bottom left: the line widths of $^{13}$CO emission line ($\Delta V_{^{13}\mathrm{CO}}$) v.s. the optical depths of $^{13}$CO emission line ($\tau_{13}$). The black line is y=x.
	Bottom right: $\tau_{13}$ integral to $\Delta V_{^{13}\mathrm{CO}}$ v.s. $\tau_{18}$ integral to $\Delta V_{\mathrm{C}^{18}\mathrm{O}}$ multiply by $\alpha_0$. $\alpha_0$ is the abundant ratio between $^{13}$CO and C$^{18}$O with the value 5.5. The black line is y=x.
	The data of IQuad and ACent cores are represented as blue dots and orange trangles, respectively.}
	\label{emi-line-para}
\end{figure*}

The probability distributions of $^{12}$CO centroid velocities (V$_{^{12}\mathrm{CO}}$) minus $^{13}$CO centroid 
velocities (V$_{^{13}\mathrm{CO}}$) in IQuad and ACent are plotted in the top left panel of Figure \ref{emi-line-para}, showing V$_{^{12}\mathrm{CO}}$ and V$_{^{13}\mathrm{CO}}$ are coincident with each other quite well. As we mentioned in Sect. \ref{subsec:profiles}, 8 blue profiles, 3 red profiles, 5 blue asymmetries and 6 red asymmetries (accounting for $\sim$15.1\%) were detected in IQuad cores and 2 blue profiles, 1 red profiles, 5 blue asymmetries and 5 red asymmetries (accounting for 13\%) in ACent cores. Only two IQuad cores are identified to have wings in the $^{12}$CO lines. The detection rate of asymmetry profiles in IQuad and ACent are simillar to that (18.2\%) found in IIQuad \citep{2016ApJS..224...43Z}, and are very low when compared with massive star forming regions such as Orion where the detection rate is as high as 70\% \citep{2008ApJ...688L..87V}. 

The blue excess E defined as (N$_\mathrm{blue}$-N$_\mathrm{red}$)/N$_\mathrm{tot}$ is $\sim$0.03 in IQuad and is 0.01 in ACent, where N$_\mathrm{blue}$ and N$_\mathrm{red}$ are the core numbers with detected blue and red profiles respectively and N$_\mathrm{tot}$ is the total number of cores \citep{1997ApJ...489..719M}. When comparing with the Class --1, 0 and I objects \citep[E$\approx$0.3; where Class --1 object is prestellar core, and protostar spends 2$\times10^4$ yr in the Class 0 phase and 2$\times10^5$ yr in the Class I phase;][]{2003cdsf.conf..157E} and the UC H\,{\footnotesize II}\ regions \citep[E=0.58;][]{2007ApJ...669L..37W}, our samples have much smaller E, which may indicate PGCC cores are still in early evolutionary stages of star formation. The star formation is not active in the PGCC cores yet.

For the suprathermal line widths, instead of part of a globle velocity structure \citep[either collapse, expansion and rotation; e.g.,][]{1974ApJ...190..557L,1992ApJ...394..204Z}, the contributions of the line widths broadening are thermal, non-thermal motions and the saturation effects \citep[the opacity broadening effect; e.g.,][]{1999ApJ...517..209G}. The opacity effect can seriously affect the optically thick $^{12}$CO emission line \citep{1979ApJ...231..720P}. The top right panel of Figure \ref{emi-line-para} shows that few sources have $\tau_{\rm ^{13}CO} \sim$ 3.0 while most of them have $\tau_{\rm ^{13}CO} <$ 1. An optical depth 3 will contribute to the line width samller than 30 percent \citep[see Equation 3 of][]{1979ApJ...231..720P}, thus the opacity broadening effect on $\Delta V_{^{13}\mathrm{CO}}$ should not be significant. It may indicate C$^{18}$O is only fully excited in the dense part of molecular cores compared with $^{13}$CO. This can also be confirmed by the comparison between the H$_2$ column densitis derived from $^{13}$CO and C$^{18}$O. The $\Delta V_{^{13}\mathrm{CO}}$ and $\Delta V_{\mathrm{C^{18}O}}$ show good correlation in both IQuad and ACent regions, with $\Delta V_{^{13}\mathrm{CO}}$ is slightly larger than $\Delta V_{\mathrm{C^{18}O}}$ (bottom left panel of Figure \ref{emi-line-para}).  From the bottom right panel of Figure \ref{emi-line-para}, it is clear that the optical depth of $^{13}$CO integrated over velocity ($\int{\tau_{13}}{dV_{^{13}CO}}$) is systematically larger than the the optical depth of C$^{18}$O integrated over velocity multiplied by $\alpha_0$ ($\int{\tau_{18}}{dV_{C{^{18}O}}}\alpha_0$). $\alpha_0$ is the abundant ratio between $^{13}$CO and C$^{18}$O with the value 5.5 \citep[and the references therein]{1983ApJ...264..517M}. There are two reasons that may result in $\int{\tau_{13}}{dV_{^{13}CO}} > \int{\tau_{18}}{dV_{C{^{18}O}}}\alpha_0$. One is that the $^{13}$CO emission line may systematically trace more molecular hydrogen than C$^{18}$O. Another reason may be the variance of x[C$^{18}$O]/x[$^{13}$CO] along the Galactocentric distance \citep{1994ARA&A..32..191W, 2013A&A...554A.103P}.

\subsection{Non-thermal motions \label{ntm}}

Analyzing non-thermal motions of dense cores is useful for understanding dynamical processes and heating transport in star formation \citep{2003RMxAC..15..293C}. 
The non-thermal motions originated from star forming activities such as infall motions and outflows can broaden the non-thermal 
velocity dispersion. However, these motions in PGCCs are not very active when compared with typical star forming regions \citep{2012ApJ...756...76W}. 
The non-thermal motions in these clumps are therefore mainly contributed by turbulent motion. 

\begin{figure*}
    \centering
    \includegraphics[width=0.3\linewidth,height=0.3\linewidth]{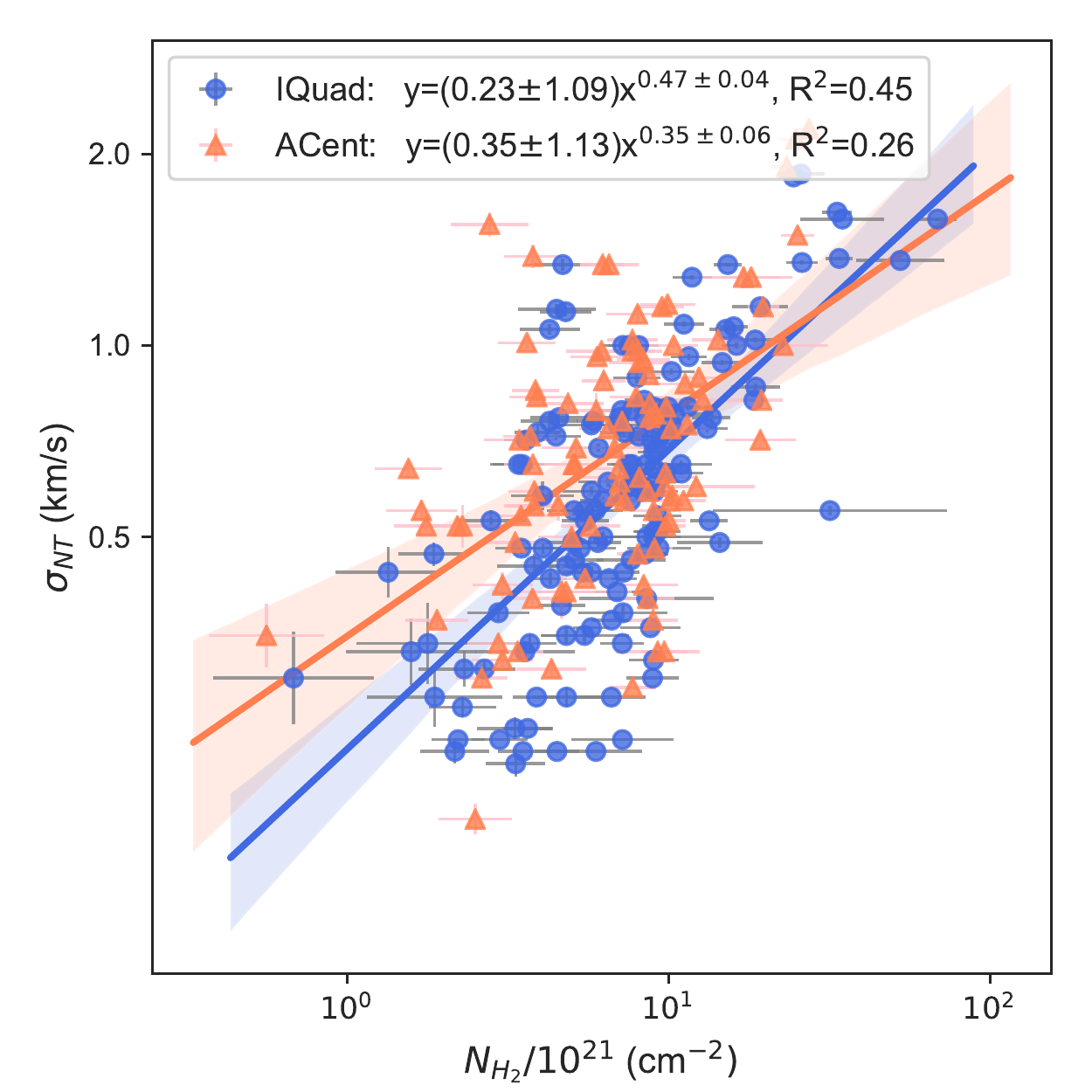}
    \includegraphics[width=0.3\linewidth,height=0.3\linewidth]{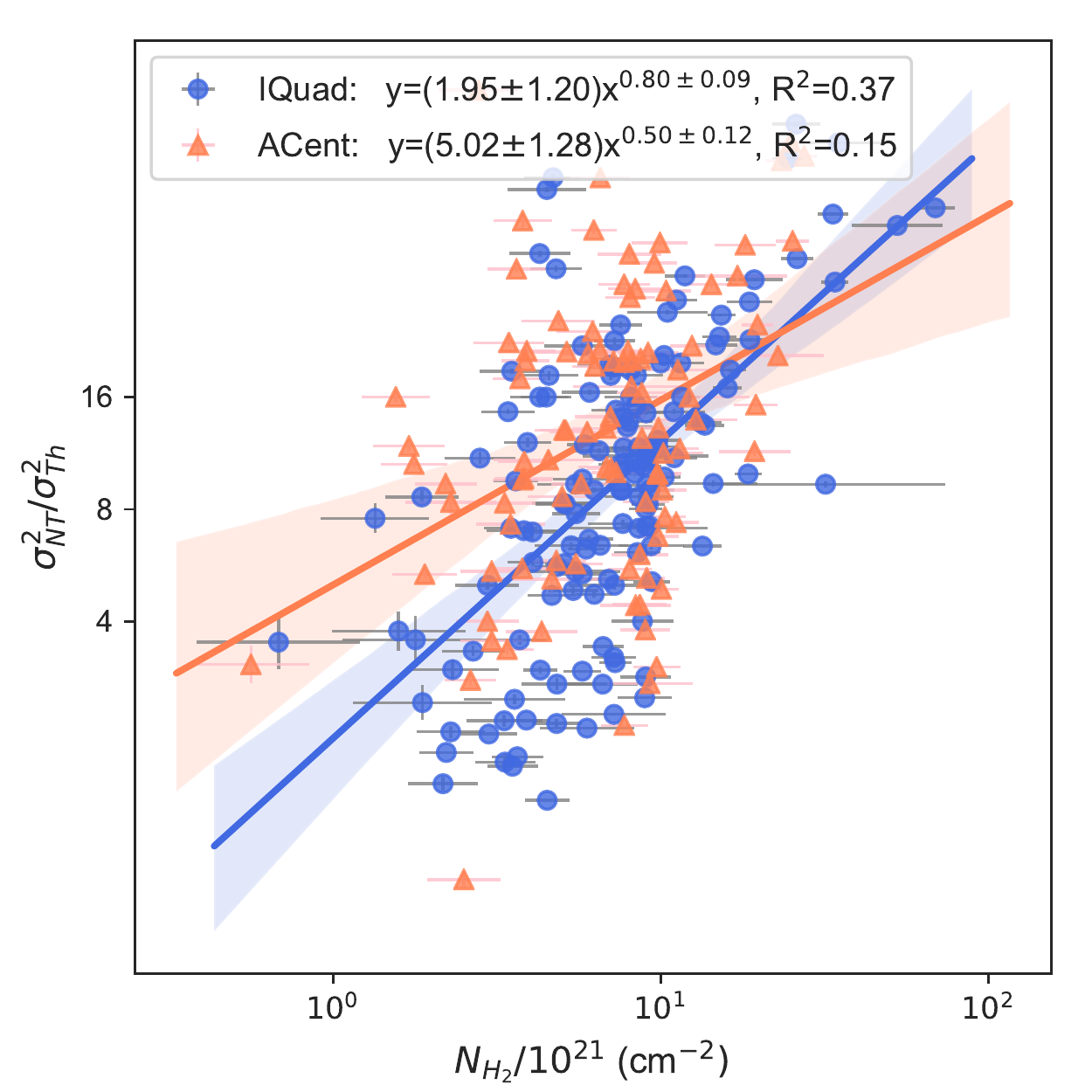}
    \includegraphics[width=0.3\linewidth,height=0.3\linewidth]{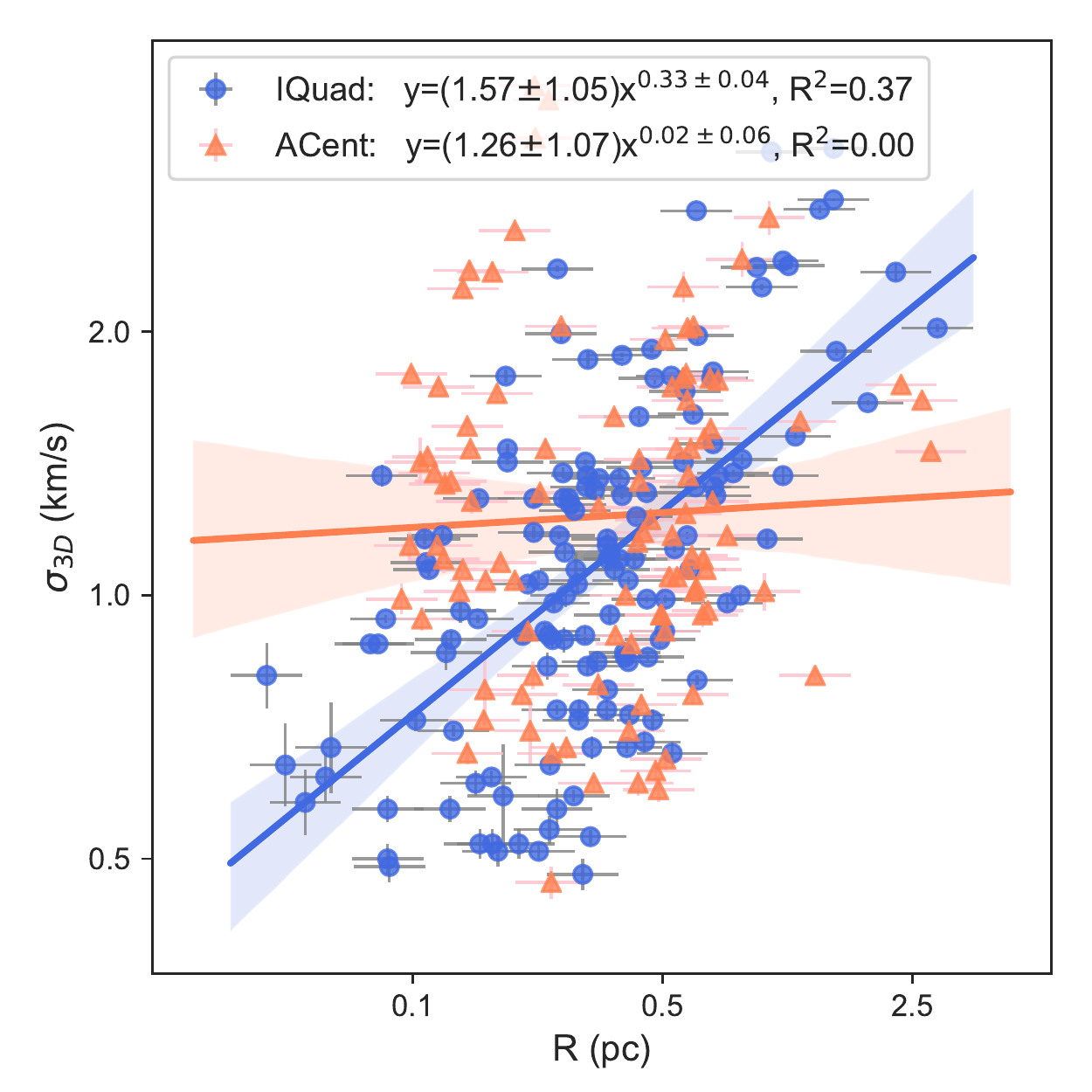}
	\caption{Left: the non-thermal velosity dispersion v.s. the column density of H$_2$.
	Middle: the ratio of non-thermal to thermal velocity dispersion v.s. the column density of H$_2$.
	Right: the Larson relationship.
	The power-law fitting is applied and their fitting results are labeled in the top of each panel. The shadow is the uncertainty of fitting results.
	The data of IQuad and ACent cores are the same as Figure \ref{emi-line-para}. }
	\label{derived-para}
\end{figure*}

The ratio of non-thermal to thermal velocity dispersion R$_{\mathrm{p}}$ = $\sigma_{\mathrm{NT}}^2/\sigma_{\mathrm{Th}}^2$ can be used to estimate the ratio of pressure contributed from the non-thermal and thermal motions \citep{2008ApJ...672..410L,2012ApJS..202....4L}. Our results show that R$_{\mathrm{p}}$ are larger than 1 for almost all sources, 
and thus turbulences in PGCC cores in both regions are typically supersonic.
The increasing trends of $\sigma_{\mathrm{NT}}$ and R$_{\mathrm{p}}$ with respect to $N_{\mathrm{H_2}}$ are found in IQuad (see the left and middle panels of Figure \ref{derived-para}). Similar relations are also found in ACent but with weak correlation. It implys that the non-thermal pressure is more dominant in denser cores. 
R$_{\mathrm{p}}$ for the IQuad cores (3.4$\pm$0.19) is on average smaller than ACent cores (3.9$\pm$0.23), suggesting that turbulence has on average dissipated more in the IQuad cores. The uncertainty of each R$_{\mathrm{p}}$ is transferred by $\sigma_{\mathrm{NT}}$ and $\sigma_{\mathrm{Th}}$.

%\subsection{The Larson Relationship}
Since \citet{1981MNRAS.194..809L} put forward the idea that the internal velocity dispersion is correlated with its region size from 0.1 to 100 pc in a power-law form, lots of works were followed \citep{2010ApJ...712.1049S, 2017MNRAS.465.1316M, 2018MNRAS.477.2220T}. We also test this correlation in IQuad and ACent (see the right panel of Figure \ref{derived-para}). For IQuad cores, the correlation can be represented as 
\begin{equation}\sigma_{3D} = (1.57\pm0.05)\times R^{0.33\pm0.04},\end{equation} with R$^2$=0.37.
The power-law index is similar to the Kolmogorov cascade ($\sigma \propto R^{1/3}$). The uncertainty of distance estimated may be a reason resulting in the weak correlation. However, the Larson relationship may be not valid on small scale, since turbulence dominates the clump structure and density distribution on large scale but not on small scale \citep{1994ApJ...423..681V}. Small scale cores are more easily affected by the density fluctuations \citep{1987A&A...172..293B}. The velocity dispertion is not related to the core size in ACent, which can confirm that turbulence does not dominate on core structure.

\subsection{PGCCs in Different Molecular Cloud Complexes} \label{different-complexes}
The internal velocity dispersion of the cloud is one of the important factors to determine the mass of the formed stars, as suggested by \citet{2001PASJ...53.1037S}. From the empirical correlation that line width is in generally found to increase with luminosity \citep[$\Delta V\propto L^{0.13-0.19}$;][]{1991ApJ...372L..95M,1999ApJS..125..161J}, \citet{2009A&A...507..369W} tested the $\Delta V-L$ relation toward red IRAS sources and suggested that the $\Delta V$ criterion refers to $\Delta V_{\rm ^{13}CO} >$ 2 km s$^{-1}$ as a characteristic value of high mass cores. We classified 10 sub-regions to check whether the cores with broad line width are massive by the line width criterion. The division of the sub-regions is from \citet{1987ApJ...322..706D} and \citet{1978ApJS...38..309H}.

When clumps are located within the (l, b) ranges of OB associations, we defined the clumps to be associated with OB associations. The limits of OB associations are from \citet{1978ApJS...38..309H}. In Figure \ref{Galactic-plane}, Serpens, Vulpecula, Cygnus, Cepheus, Auriga, Gemini, Monoceros and Canis Major each hosts at least one OB association. In Vulpecula, Cygnus, Cepheus, Monoceros and Canis Major, some of clumps are located in OB associations. The statistics of $\Delta V_{\rm ^{13}CO}$, $\sigma_{3\mathrm{D}}$, T$_{\rm ex}$, N$_{\rm H_2}$ and M in these sub-regions are listed in Table \ref{stat-complexes}.

In Table \ref{stat-complexes}, $\Delta V_{\rm ^{13}CO}$ of the cores in Canis Major are averaged larger than 2 km s$^{-1}$, but the average core mass in Canis Major is smaller than in every other sub-regions apart from Aquila. Additionally, 
PGCC cores towards Serpens, Cloud A/B, Vulpecula, Cepheus and Auriga regions have narrow line widths with average values closed to 1.3 km s$^{-1}$ \citep[the typical line width of low mass cores;][]{1983ApJ...264..517M} and high masses with average values larger than 8 M$_{\bigodot}$. These results indicate that the line width criterion seems to be broken down in our sample. Low mass cores with broad line width may be still turbulent \citep{2006PASJ...58..343S}. Fragmentation are suggested in some massive cores with narrow line width.

\subsection{Core status}
\subsubsection{Core mass function}
\begin{figure*}
	\includegraphics[width=0.33\linewidth]{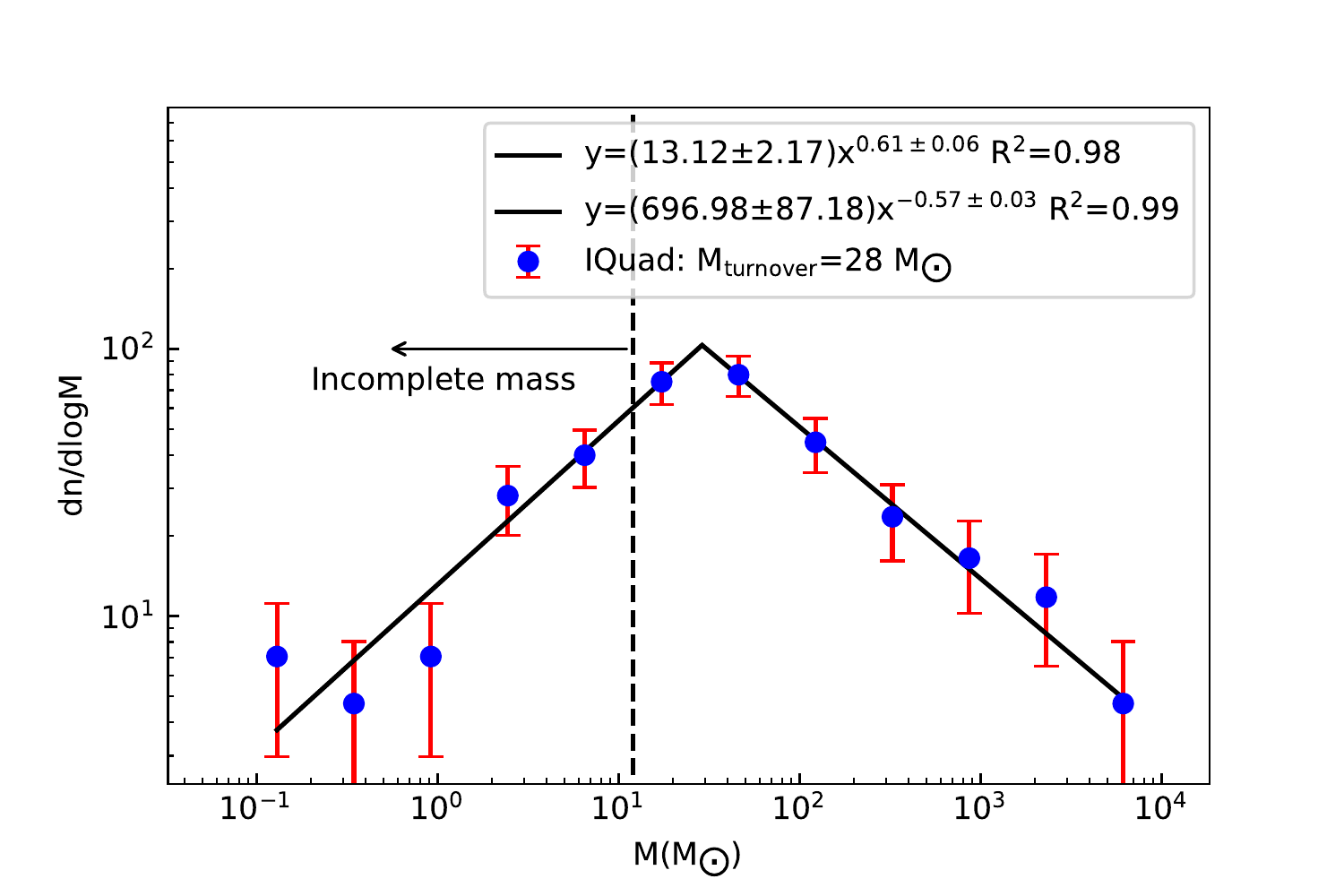}
	\includegraphics[width=0.33\linewidth]{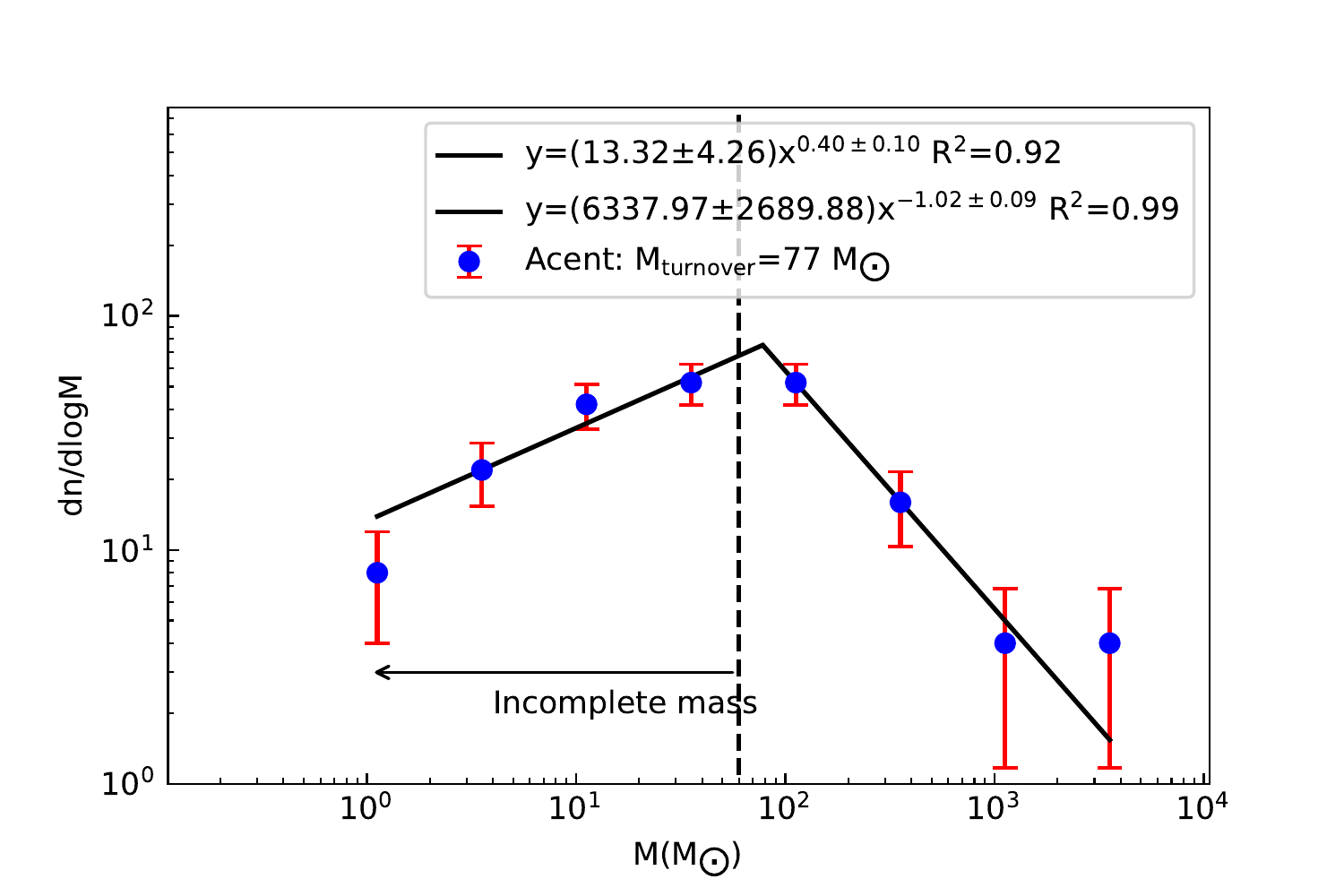}
	\includegraphics[width=0.33\linewidth]{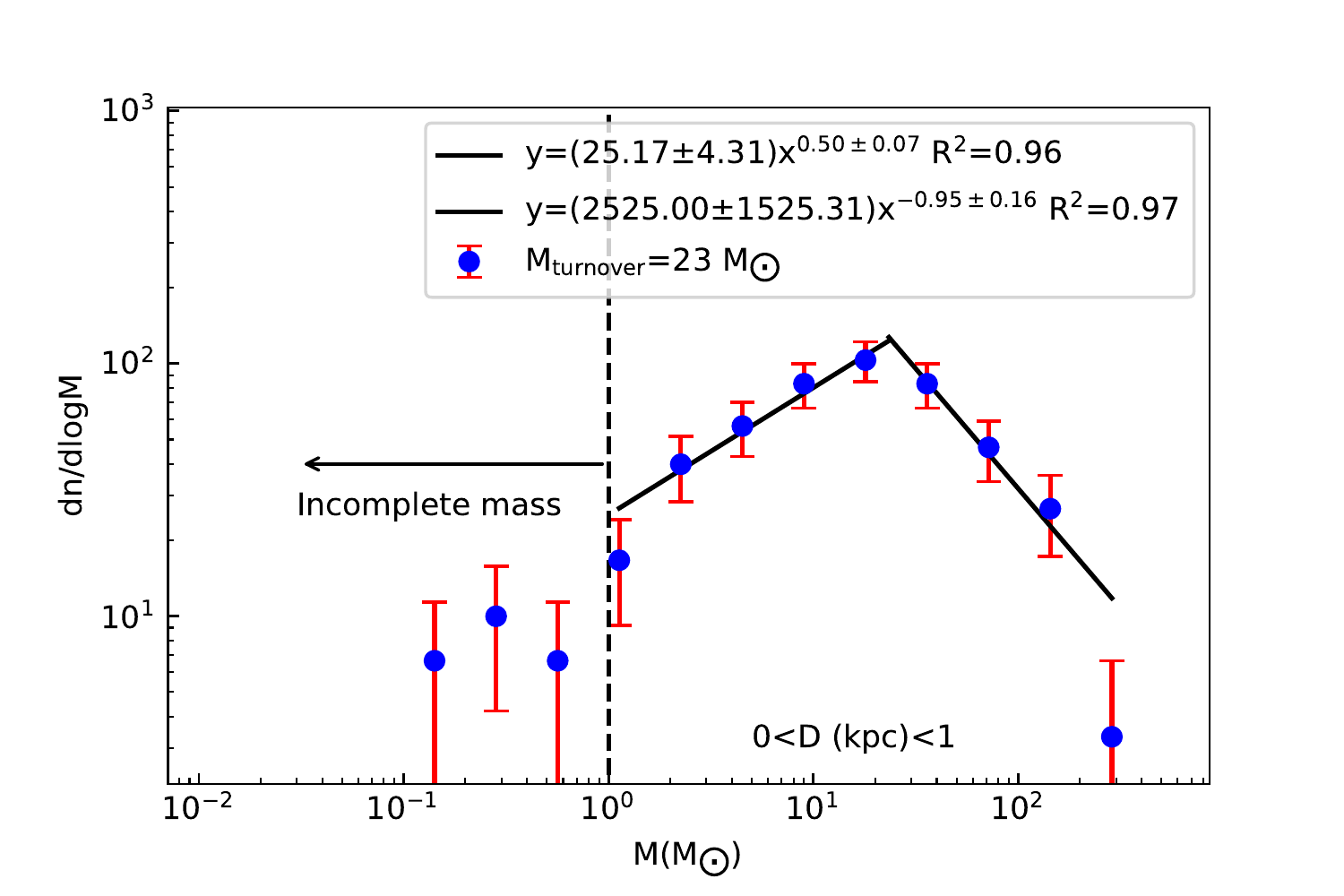}
	\caption{The core mass function: Left and middle panels are plotted for the cores in IQuad and ACent respectively. Right panel is plotted for the cores with distance less than 1 kpc. The error bars correspond to statistical uncertainties of $\sqrt{N}$. The black dashed line is plotted as the mass completeness limit.
	}
	\label{CMF}
\end{figure*}

The core mass function (CMF) is usually found to follow $dN/dlogM \sim M^{-\alpha_{CMF}}$. The CMFs in the two regions are plotted in Figure \ref{CMF}. The left and middle panels present the CMF for the cores in IQuad and ACent respectively. Based on Equation (1), (2), (4) and (9), the mass is proportional to the brigtness temperature (T$_{\rm b}$) of $^{12}$CO, integrated intensity of $^{13}$CO, the optical depth of $^{13}$CO ($\tau_{13}$), source size in arc ($\frac{1}{2}\sqrt{ab}$) and distance (D). Most of the PGCC cores in this paper should have the median values of these parameters. However, the dense cores, we focused, have an isoline of 50\% $^{13}$CO peak intensity. So, we adopted half of the median value of the peak $^{13}$CO integrated intensity (2.8 K km s$^{-1}$), the median values of T$_{\rm b}$($^{12}$CO) (8.1 K), $\tau_{13}$ (0.6), source size (2.$'$5) and the largest value of distance in each region (4.15 kpc in IQuad and 9.17 kpc in ACent) to calculate the mass completeness limit. The mass completeness limits in IQuad and ACent are 12 M$_{\bigodot}$ and 60 M$_{\bigodot}$ respectively. Power-law fittings are found at the high and low mass end of CMFs. The shape of the CMF may be affected by incompleteness limits in the low mass end. The fitting results give y = (696.98$\pm$87.18)x$^{-0.57\pm0.03}$ with R$^2$=0.99 in IQuad and y = (6337.97$\pm$11.93)x$^{-1.02\pm0.04}$ with R$^2$=0.99 in ACent. The $\alpha_{\rm CMF}$ values for IQuad and ACent are 0.57$\pm$0.03 and 1.02$\pm$0.04 respectively. 
The slopes change at 28 M$_{\bigodot}$ (the turnover mass) in IQuad and 77 M$_{\bigodot}$ in ACent. The turnover mass in ACent is similar to the mass completeness limit and few data points can be used to fit as the power-law function in the high mass end. So the turnover mass and $\alpha_{\rm CMF}$ in ACent are unreliable.

It is important to make comparison of CMF to stars initial mass function (IMF) which is originally derived by \citep{1955ApJ...121..161S}, since IMF is related to CMF \citep{2007MNRAS.374.1413N,2007PASP..119..855W}. Based on 1.3 mm continuum observations of $\rho$ Ophiuchi, \citet{1998A&A...336..150M} derived CMF with an index of $\sim$1.5 for core mass $\geq 1 M_{\bigodot}$ and $\sim$0.5 for core mass $\leq 1 M_{\bigodot}$. The derived CMF in $\rho$ Ophiuchi has a similar power-law index to IMF, suggesting the star mass is related to core mass. Similar conclusion was reached when studying Serpens \citep{1998ApJ...508L..91T} and Aquila Rift \citep{2010A&A...518L.106K}. The $\alpha_{\rm CMF}$ of 0.57 for IQuad cores is flatten when comparing with the common IMF for stars with M $> M_{\bigodot}$ \citep[$\sim$1.5; e.g.,][]{1974AJ.....79.1280T,1980MNRAS.191..511H,1984A&A...133...99C}. The distant cores with high-mass may have fragmentation and may harbor multiple sub-components which can not be resolved in this low spatial resolution observations. This can explain the flatten slope of CMF found in IQuad. The $\alpha_{\rm CMF}$ of 0.95 for cores with distance less than 1 kpc (see right panel of Figure \ref{CMF}) is significantly larger than 0.57 for IQuad cores, which may support that the fragmentation exists in the distant cores. Alternatively, turbelent pressure and magnetic filed tension may decrease the efficiency of rising core mass, which can explain the flatten slope found in CO gas cores \citep{2011IAUS..270..255A}.

Comparison the turnover mass of CMF M$_{\rm turnover}$(CMF) to that of IMF M$_{\rm turnover}$(IMF) is useful for understanding the formation process of gas cores \citep{2015A&A...584A..91K}. Star formation efficiency could be roughly estimated by M$_{\rm turnover}$(IMF) devided by M$_{\rm turnover}$(CMF) \citep{1998A&A...336..150M,2007A&A...462L..17A,2007MNRAS.374.1413N}. M$_{\rm turnover}$(IMF) is given about 0.25 M$_{\bigodot}$ for Cygnus-X with a slope of $\sim$0.7 \citep{2016MNRAS.458.3027M}. We take the value of 0.25 M$_{\bigodot}$ as M$_{\rm turnover}$(IMF) to calculate the star formation efficiency. The star formation efficiency is $\sim$0.9\% for IQuad cores. For the cores with distance less than 1 kpc in the two regions, the star formation efficiency is $\sim$1.1\%. The star formation efficiencies found here are significantly lower than $\sim$3\%--6\% for common star formation regions \citep{2009ApJS..181..321E}, suggesting that PGCC cores are in the early evolutionary stage of star formation and whether the cores can form stars is yet to be decided. The argument that the distant cores may have unresolved fragmentation in the preceding paragraph seems to suggest that the more distant objects are clumps, rather than individual star-forming cores. This might also be an additional or alternative explanation for the apparent low star formation efficiency of the PGCC cores.

\subsubsection{Gravitational stabilities of the dense cores} \label{gs}
The virial parameter $\alpha$ = M$_{\mathrm{vir}}$/M, shown in column (9) of Table \ref{area-para}, can be applied to discribe the stabilities of gas cores \citep{2013ApJ...779..185K}. We defined that virial mass is consistent with a mass within a factor of 3 \citep{2012ApJS..202....4L}. In IQuad, one-hundred-three (70.5\%) dense cores have virial masses consistent with the core masses. This ratio is larger than that in ACent (56\%), but smaller than that in Orion complex \citep[88\%;][]{2012ApJS..202....4L}. About 58\% (85) IQuad cores and 45\% (45) ACent cores have Jeans masses consistent with masses within a factor of 3, i.e. the other cores have Jeans instability and will be fragmentations or collapses \citep{2008PASJ...60..421F}.

\begin{figure*}
	\includegraphics[width=0.45\linewidth]{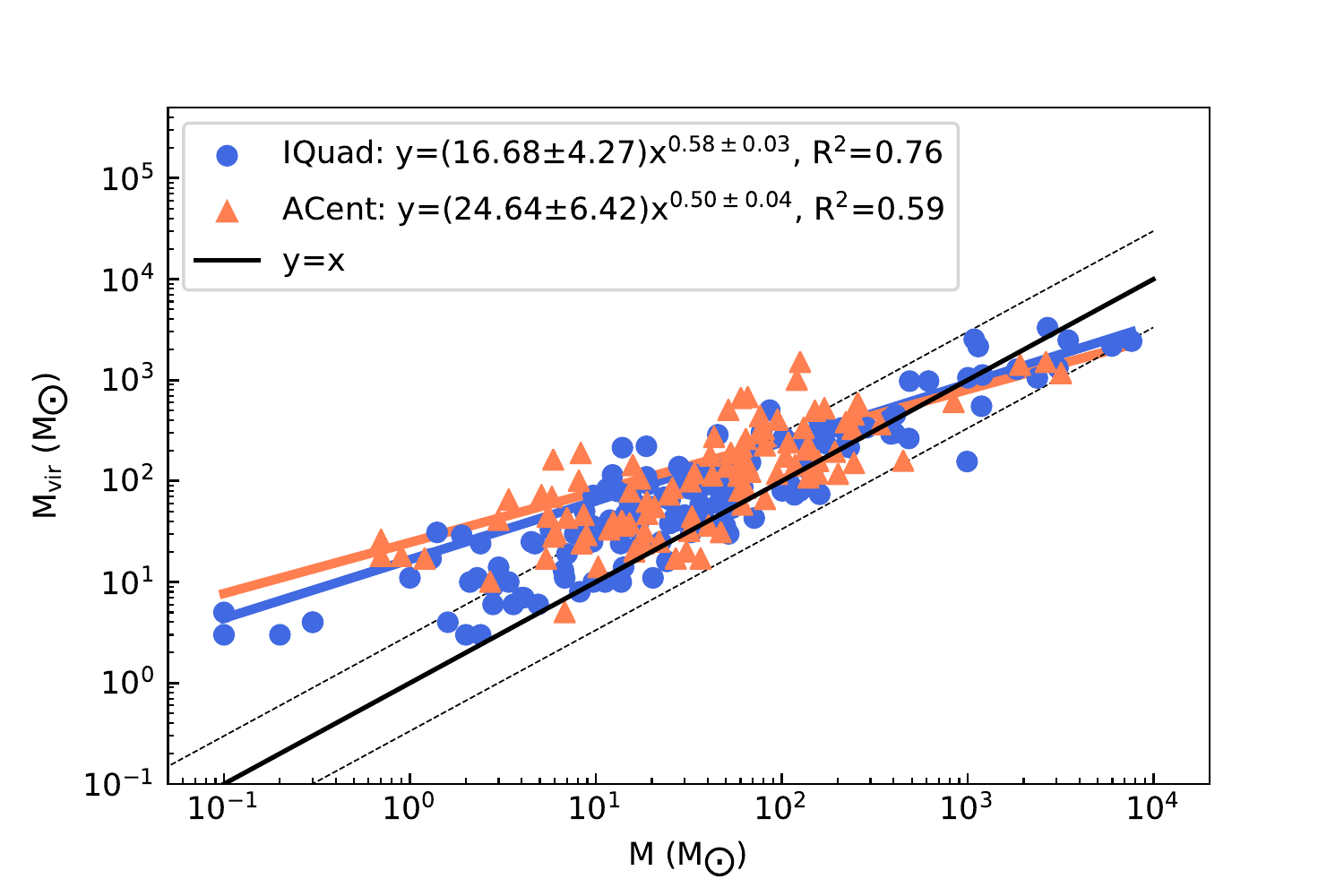}
	\includegraphics[width=0.45\linewidth]{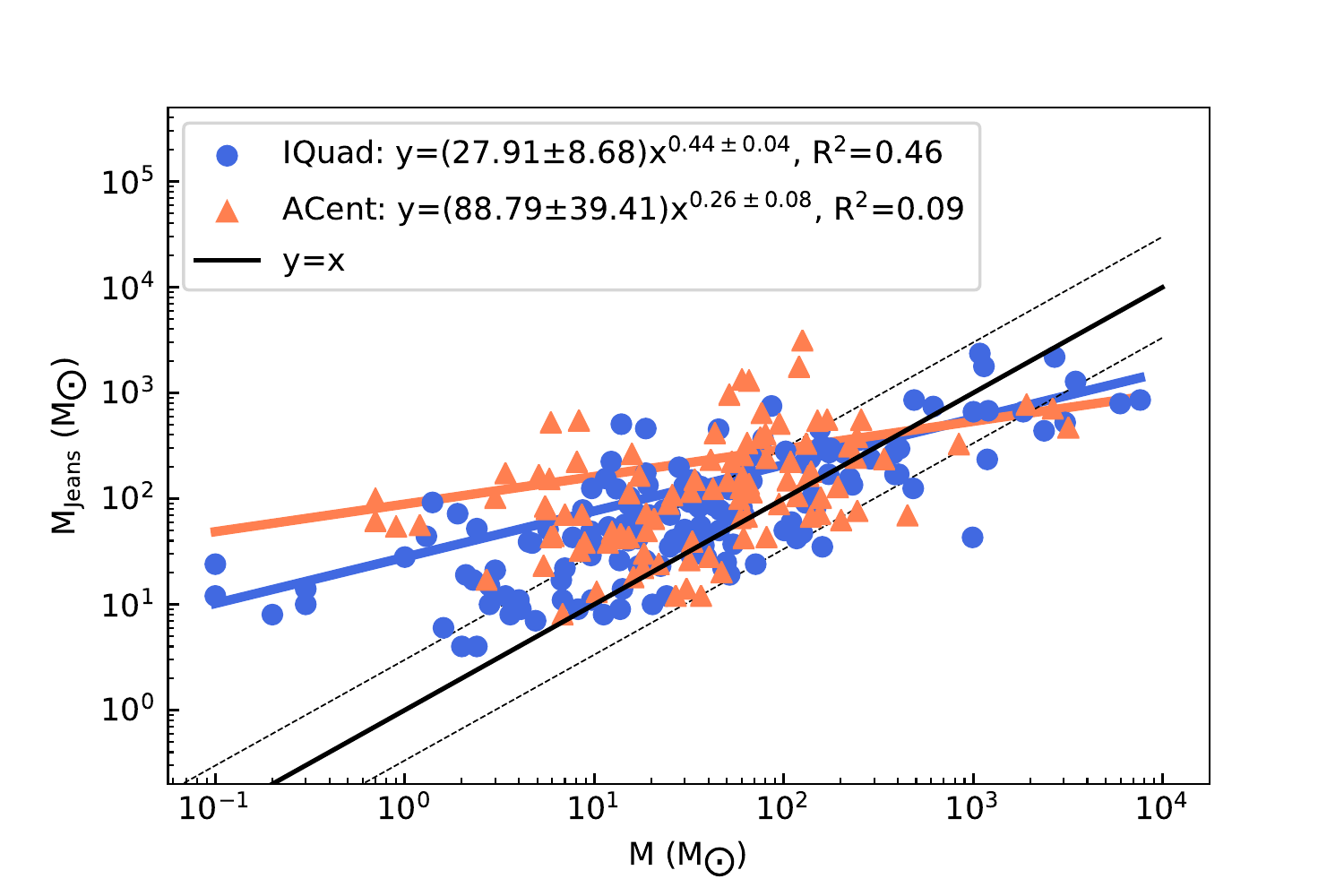}
	\caption{Left panel: plots of virial mass v.s. gas mass. Right panel: plots of Jeans mass v.s. gas mass.
	The solid black line is plotted as y=x, while two dashed black lines are plotted as virial mass consistent with mass within a factor of 3. The data of IQuad and ACent cores are the same as Figure \ref{emi-line-para}. }
	\label{Stability-analyzing}
\end{figure*}

The relationships between M$_{\mathrm{vir}}$ and M in both regions are shown in the left panel of Figure \ref{Stability-analyzing}. The pawer-low fitting gives a power index of 0.58$\pm$0.03 and 0.50$\pm$0.04 for IQuad and ACent, respectively. According to Equation (9) and (10), the power-law indexes of $\sim$0.5 may indicate that the relationship between M$_{\mathrm{vir}}$ and M is almost entirely dependent on R and thus the velocity dispersion between cores has random variation. The power-law indexes for the cores obtained here are significantly larger than $\frac{1}{3}$ for pressure-confined cores \citep{1992ApJ...395..140B}, indicating that the cores are most likely not pressure-confined, but gravitationally bound \citep{2009ApJ...691.1560I}, unless the pressures are significantly supported by magnetic fields \citep{2013ApJ...779..185K}.

Most of the cores have M$_{\mathrm{Jeans}}$ larger than M (see right panel of Figure \ref{Stability-analyzing}). In fact, 104 IQuad cores and 78 ACent cores have M$_{\mathrm{Jeans}}$ larger than M. Two cores of G093.51-04.31C1 and G181.71+04.16C2 having M$_{\mathrm{Jeans}}<$ M show blue profile, which indicate that they may be gravitationally bound and tend to collapse.

\subsubsection{Associated objects}
Matching gas cores with stellar objects is very useful for understanding the environment and evolutional status of the cloud cores. Based on the AllWISE catalogue released by the \emph{Wide-field Infrared Survey Explorer} (WISE) mission, \citet{2016MNRAS.458.3479M} presented a catalogue of identified 133 980 Class I/II YSO candidates (CIs) which have a significant infrared (IR) excess. The good correlation between the surface density distribution of CIs and that of clumps is found in the Taurus-Auriga-Perseus-California region \citep{2016MNRAS.458.3479M}. We have matched clumps with the CIs from this catalogue, and the Class III YSO candidates (CIIIs) from the CIIIs catalogue \citep{2016MNRAS.458.3479M}, and \emph{IRAS} point source (IR-source) catalogue \citep{1988SSSC..C......0H}. The CIIIs have IR excess and its IR color similar to main sequence stars. The CIs, CIIIs and IR-sources are believed to be associated with clumps if they are located in the mapped area of 14$'\times$14$'$. About 61\% of IQuad clumps (46) and 59\% of ACent clumps (29) are associated with CIs, indicating that CIs have a good correlation in spatial distribution with clumps in IQuad and ACent. About 95\% of IQuad clumps (72) and 80\% of ACent clumps (39) are associated with the objects of CIs, CIIIs and/or IR-sources.

The objects are believed to be associated with the cores if it has an offset to the core center less than a beam size ($\sim$1\arcmin) and the distances of associated objects (if we can get from the catalogues) similar to the cores distances. We found that $\sim$8.9\% (13) of IQuad cores are associated with CIs, 18\% (18) of ACent cores are associated with CIs excluding 2 cores which have associated CIs but the difference of core distance and CIs distance is very large. In totall, 31 of 246 cores ($\sim$12.6\%) are associated with CIs, which can not indicate that there is a good correlation between the CIs and the cores in this work. In total, 58 cores ($\sim$23.6\%) are identified to be associated with the objects. Out of the 58 cores, 3 cores of G028.45-06.39C2, G095.51+09.97aC1 and G226.80-07.04C1 have signs of star formation in the $^{12}$CO line, indicating that the three cores are more evolved than other cores. All these PGCC cores with associated objects are listed in Table \ref{association}.

\section{Summary}
We have performed a mapping survey in J=1-0 transition of $^{12}$CO, $^{13}$CO and C$^{18}$O with PMO 13.7-m telescope toward 104 Planck Galactic cold clumps (PGCCs). Among the mapped sources, 65 are located in the First Quadrant (IQuad) and 39 in the Anti-Center direction region (ACent). $^{12}$CO and $^{13}$CO lines were detected in all of them, while C$^{18}$O line was detected in 53 of IQuad PGCCs with the detection rate about 81.5\% and was detected in 31 of ACent PGCCs with the detection rate about 79.5\%. One hundred and twenty five velocity components (hereafter clumps) are identified in total, including 76 in IQuad and 49 in ACent. We identified 246 dense cores from the velocity integrated intensity maps, with 146 cores in IQuad and 100 ones in ACent. 
The parameters of detected lines and identified cores were derived. We have discussed their properties, morphologies and compared the differences of two regions. The main findings are summarized as follows.

\begin{enumerate}
	\item Our samples are mostly located on the edge of the CO gas detected in Columbia 1.2 telescope, indicating additional gas was detected.
	\item The excitation temperature (T$_{\rm ex}$) of IQuad cores and ACent cores ranges from 7.6 to 22.5 K and 7.2 to 21.8 K respectively. The excitation temperature is similar to the dust temperature (T$_{\rm dust}$) obtained in \citet{2016A&A...594A..28P}. However, the median value of the T$_{\rm ex}$ is smaller than that of T$_{\rm dust}$, suggesting that gas and dust are well coupled in these cores and gas might be heated by dust.
	\item Non-thermal velocity dispersion and the non-thermal to thermal pressure increases with H$_2$ column density indicates that the non-thermal pressure is more dominant in denser cores. The correlation of the velocity dispersion versus core size is weak in IQuad and the velocity dispersion is not related to core size in ACent, suggesting that Larson relationship seems not to be valid in PGCC cores in IQuad and ACent.
	\item Core mass function (CMF) fittings give the power-law indexes --0.56 for IQuad cores, --1.02 for ACent cores. The flat slope may be caused by that the fragmentations exist in the distant cores. The turnover mass of the CMF is 28 M$_{\bigodot}$ in IQuad and 77 M$_{\bigodot}$ in ACent. The mass completeness limit is 12 M$_{\bigodot}$ in IQuad and 60 M$_{\bigodot}$ in ACent. The similarity between mass completeness limit and the turnover mass is found in ACent, resulting in that turnover mass in ACent is unreliable. The lower star formation efficiency of $\sim$0.9\% is found in IQuad when we compare the turnover mass of CMF and that of IMF, indicating that PGCC cores are in the early evolutionary stage of star formation.
	\item The correlation of M$_{\mathrm{vir}}$ versus M shows the power-law index of 0.58 for IQuad cores and 0.50 for ACent cores which are significantly larger than $\frac{1}{3}$ for pressure-confined cores \citep{1992ApJ...395..140B}, indicating that most of the cores are not likely pressure-confined, but gravitationally bound. 
	\item The non-Gaussian profiles, including 8 blue profiles, 3 red profiles and 2 wings, are detected in IQuad cores, while 2 blue profiles and 1 red profiles are identified in ACent cores. The lower blue excess of $\sim$0.03 in IQuad and 0.01 in ACent suggestes that the star formation us not active in PGCC cores. We have identified 58 cores to be associated with Class I/II YSO candidates, Class III YSO candidates and/or IRAS pointing sources, 34 of them are located in IQuad and 24 of them in ACent. Out of the 58 cores, three cores have signs of star formation in the detected $^{12}$CO lines and may be more evolved than other cores.

\end{enumerate}

We are grateful to the staff at the Qinghai Station of PMO for their assistance during the observations. The Wisconsin H$\alpha$ Mapper and its H$\alpha$ Sky Survey have been funded primarily by the National Science Foundation. The facility was designed and built with the help of the University of Wisconsin Graduate School, Physical Sciences Lab, and Space Astronomy Lab. NOAO staff at Kitt Peak and Cerro Tololo provided on-site support for its remote operation. This work was supported by the NSFC No. 11988101, NSFC Nos. 11433008, 11373009, 11503035, 11573036 and U1631237, and China Ministry of Science and Technology under State Key Development Program for Basic Research (No. 2012CB821800), and the Top Talents Program of Yunnan Province (2015HA030).

\bibliographystyle{aasjournal}

      \bibliography{AAS19777}

\clearpage
\begin{deluxetable*}{ccccc cccc}
\tablecaption{Sample of PGCCs
\label{samples}}
\tablewidth{0pt}
\tablehead{
\colhead{Name} & \colhead{Glon} & \colhead{Glat} & \colhead{R.A.(J2000)} &
\colhead{Decl.(J2000)} & \colhead{R.A.(B1950)} & \colhead{Decl.(B1950)} & \colhead{Sub-regions\tablenotemark{\,{\footnotesize a}}} & \colhead{OB association\tablenotemark{\,{\footnotesize b}}}\\
\colhead{} & \colhead{deg} & \colhead{deg} & \colhead{h:m:s} &
\colhead{d:m:s} & \colhead{h:m:s} & \colhead{d:m:s} & \colhead{} & \colhead{}
}
\decimalcolnumbers
\startdata
\multicolumn{9}{c}{The First Galactic Quadrant (IQuad)}\\
\hline
G013.86+04.55  &  13.864745  &  4.5556397  &  17:59:03.17  &  -14:41:26.77  &  17:56:11.93  &  -14:41:16.39  &  Serpens & \\
G014.74+04.06  &  14.743651  &  4.0693364  &  18:02:34.88  &  -14:09:59.84  &  17:59:44.29  &  -14:10:04.91  &  Serpens & \\
G015.16+07.23  &  15.161132  &  7.2371593  &  17:52:06.35  &  -12:14:27.93  &  17:49:18.15  &  -12:13:47.29  &  Serpens & \\
G026.45+08.02  &  26.455076  &  8.0275078  &  18:11:02.76  &  -02:02:51.60  &  18:08:26.64  &  -02:03:34.20  &  Aquila & \\
G028.45-06.39  &  28.454588  &  -6.3918767  &  19:06:09.21  &  -06:52:51.78  &  19:03:27.70  &  -06:57:31.44  &  Aquila & \\
G031.26-05.37  &  31.267088  &  -5.3793736  &  19:07:35.28  &  -03:55:38.37  &  19:04:57.11  &  -04:00:24.16  &  Aquila & \\
G034.69-06.57  &  34.694824  &  -6.5795875  &  19:18:04.25  &  -01:26:07.30  &  19:15:28.89  &  -01:31:36.81  &  Aquila & \\
G038.36-00.95  &  38.364254  &  -0.95123816  &  19:04:45.73  &  +04:23:50.61  &  19:02:16.91  &  +04:19:16.36  &  Aquila & \\
G058.97-01.66 & 58.974606 & -1.6601636 & 19:47:54.71 & +22:10:14.34 & 19:45:45.23 & +22:02:43.88 & Vulpecula &Vul OB4\\
G060.75-01.23 & 60.754391 & -1.2310537 & 19:50:13.23 & +23:55:19.72 & 19:48:05.76 & +23:47:40.19 & Vulpecula & Vul OB1\\
\hline
\multicolumn{9}{c}{The Galactic Anti-Center Direction Region (ACent)}\\
\hline
G178.48-06.76  &  178.48387  &  -6.7673697  &  05:16:16.57  &  +26:29:58.11  &  05:13:10.22  &  +26:26:41.39  &  Auriga & \\
G178.72-07.01  &  178.72557  &  -7.0115962  &  05:15:59.71  &  +26:09:46.55  &  05:12:53.85  &  +26:06:28.65  &  Auriga & \\
G178.98-06.74  &  178.98924  &  -6.7485881  &  05:17:37.38  &  +26:05:53.18  &  05:14:31.56  &  +26:02:42.26  &  Auriga & \\
G179.14-06.27  &  179.14305  &  -6.279283  &  05:19:43.91  &  +26:14:18.94  &  05:16:37.83  &  +26:11:17.07  &  Auriga & \\
G180.92+04.53  &  180.92284  &  4.5369301  &  06:05:48.75  &  +30:25:11.40  &  06:02:35.79  &  +30:25:29.78  &  Auriga & \\
G181.16+04.33  &  181.16454  &  4.3311534  &  06:05:31.15  &  +30:06:33.65  &  06:02:18.68  &  +30:06:50.77  &  Auriga & \\
G181.42-03.73  &  181.42821  &  -3.7328291  &  05:34:46.40  &  +25:44:48.16  &  05:31:40.70  &  +25:42:51.33  &  Auriga & \\
G181.71+04.16  &  181.71385  &  4.1628323  &  06:06:03.83  &  +29:32:54.89  &  06:02:52.23  &  +29:33:14.43  &  Auriga & \\
G181.84+00.31  &  181.84569  &  0.31707302  &  05:51:10.62  &  +27:31:08.15  &  05:48:02.13  &  +27:30:22.71  &  Auriga & \\
G181.88+04.49  &  181.88963  &  4.4995117  &  06:07:48.53  &  +29:33:27.39  &  06:04:36.92  &  +29:33:54.55  &  Auriga & \\
\enddata
\tablenotetext{a}{The limits of the sub-regions are from \citet{1987ApJ...322..706D} and \citet{1978ApJS...38..309H}.}
\tablenotetext{b}{The limits of the OB associations are from \citet{1978ApJS...38..309H}.}
\tablecomments{The complete table is available in machine-readable form.}
\end{deluxetable*}

\begin{deluxetable*}{ccccc ccccc ccc}
\renewcommand{\thetable}{\arabic{table}A}
\centering
\tablecaption{Line parameters of core centers
\label{obs-para}}
\tabletypesize{\scriptsize}
\tablehead{
\colhead{source} & \colhead{D} & \colhead{ T$_b$($^{12}$CO)} &
\colhead{V$_{lsr}$($^{12}$CO)} & \colhead{${\rm \Delta V_{^{12}CO}}$ } &
\colhead{T$_b$($^{13}$CO)} & \colhead{ V$_{lsr}$($^{13}$CO)} &
\colhead{${\rm \Delta V_{^{13}CO}}$} & \colhead{T$_b$(C$^{18}$O)} &
\colhead{V$_{lsr}$(C$^{18}$O)} & \colhead{${\rm \Delta V_{C^{18}O}}$} & \colhead{V$_{\mathrm{peak}}$($^{12}$CO)} & \colhead{profile\tablenotemark{\,{\footnotesize a}}} \\
\colhead{} & \colhead{(kpc)} & \colhead{$K$} & \colhead{(km\ s$^{-1}$)} & \colhead{(km\ s$^{-1}$)} &
\colhead{$K$} & \colhead{(km\ s$^{-1}$)} & \colhead{(km\ s$^{-1}$)} &
\colhead{$K$} & \colhead{(km\ s$^{-1}$)} & \colhead{(km\ s$^{-1}$)} & \colhead{(km\ s$^{-1}$)} & \colhead{}
}
\decimalcolnumbers
\startdata
\multicolumn{13}{c}{The First Galactic Quadrant (IQuad)}\\
\hline
G013.86+04.55C1  & 0.29 & 6.98(0.30) & 8.84(0.03) & 2.06(0.07) & 2.37(0.18) & 8.91(0.04) & 1.81(0.10) & \nodata & \nodata & \nodata                 &  9.18 & \nodata \\
G013.86+04.55C2  & \nodata & 8.19(0.26) & 8.92(0.02) & 2.10(0.05) & 4.48(0.18) & 8.97(0.02) & 1.43(0.04) & 0.94(0.20) & 9.00(0.10) & 1.39(0.22)     &  8.71 & \nodata \\
G013.86+04.55C3  & \nodata & 7.34(0.23) & 9.05(0.02) & 2.36(0.05) & 4.44(0.15) & 9.31(0.01) & 1.39(0.04) & 0.98(0.17) & 9.35(0.07) & 1.11(0.14)     &  9.18 & \nodata \\
G013.86+04.55C4  & \nodata & 7.78(0.38) & 9.70(0.03) & 2.87(0.11) & 3.89(0.16) & 9.49(0.02) & 1.69(0.05) & 0.82(0.17) & 9.44(0.09) & 1.13(0.17)     &  9.98 & \nodata \\
G013.86+04.55C5  & \nodata & 8.31(0.49) & 10.03(0.03) & 2.31(0.09) & 4.34(0.20) & 10.15(0.02) & 1.54(0.05) & 1.56(0.25) & 9.93(0.04) & 0.74(0.09)   & 10.13 & \nodata \\
G014.74+04.06C1  & 1.08 & 6.37(0.24) & 28.71(0.03) & 3.52(0.11) & 3.13(0.12) & 28.64(0.02) & 1.64(0.05) & 0.96(0.16) & 28.60(0.05) & 1.00(0.13)     & 28.26 & \nodata \\
G014.74+04.06C2  & \nodata & 10.89(0.25) & 29.20(0.00) & 1.95(0.03) & 4.69(0.09) & 29.07(0.01) & 1.51(0.02) & 1.00(0.10) & 29.01(0.04) & 1.22(0.09) & 29.04 & \nodata \\
G015.16+07.23aC1 & 0.25 & 9.09(0.19) & 3.10(0.01) & 1.47(0.02) & 5.36(0.15) & 2.99(0.01) & 1.21(0.02) & 0.48(0.14) & 2.82(0.09) & 0.98(0.22)        &  3.33 & \nodata \\
G015.16+07.23bC1 & 0.23 & 6.98(1.57) & 6.41(0.07) & 1.21(0.21) & 2.55(0.61) & 6.38(0.04) & 0.68(0.13) & \nodata & \nodata & \nodata                 &  6.51 & \nodata \\
G015.16+07.23bC2 & \nodata & 5.14(0.53) & 6.69(0.06) & 1.64(0.12) & 1.31(0.28) & 6.58(0.06) & 1.05(0.18) & \nodata & \nodata & \nodata              &  6.35 & \nodata \\
\hline
\multicolumn{13}{c}{The Galactic Anti-Center Direction Region (ACent)}\\
\hline
G178.48-06.76aC1 & 0.64 & 9.45(0.68) & 6.96(0.02) & 0.90(0.04) & 4.50(0.59) & 6.87(0.01) & 0.44(0.03) & 1.12(0.30) & 6.89(0.03) & 0.31(0.06)        &  6.89 & \nodata \\
G178.48-06.76aC2 & \nodata & 9.76(1.01) & 6.96(0.04) & 0.91(0.05) & 0.68(0.20) & 6.97(0.10) & 0.84(0.17) & \nodata & \nodata & \nodata              &  6.96 & \nodata \\
G178.48-06.76aC3 & \nodata & 9.91(1.28) & 7.19(0.06) & 1.18(0.11) & 7.02(0.30) & 7.09(0.01) & 0.80(0.03) & 1.39(0.31) & 7.02(0.05) & 0.56(0.09)     &  7.19 & \nodata \\
G178.48-06.76bC1 & 0.64 & 9.54(0.82) & 8.00(0.01) & 0.65(0.04) & 5.84(0.47) & 7.48(0.03) & 0.99(0.06) & 0.92(0.27) & 7.34(0.08) & 0.80(0.17)        &  8.00 & \nodata \\
G178.72-07.01C1  & 0.63 & 9.87(0.36) & 7.12(0.01) & 0.83(0.02) & 6.90(0.22) & 7.10(0.01) & 0.69(0.02) & 1.38(0.30) & 6.87(0.03) & 0.38(0.06)        &  7.31 & \nodata \\
G178.98-06.74C1  & 0.59 & 8.01(0.30) & 7.46(0.02) & 1.40(0.04) & 5.79(0.22) & 7.60(0.01) & 0.94(0.03) & 1.02(0.32) & 7.67(0.07) & 0.74(0.19)        &  7.53 & \nodata \\
G178.98-06.74C2  & \nodata & 8.48(0.28) & 7.55(0.01) & 1.26(0.03) & 6.23(0.18) & 7.70(0.01) & 0.88(0.02) & 1.86(0.27) & 7.74(0.02) & 0.43(0.05)     &  7.53 & \nodata \\
G178.98-06.74C3  & \nodata & 8.75(0.22) & 7.68(0.01) & 1.05(0.02) & 6.87(0.16) & 7.75(0.01) & 0.79(0.01) & 2.90(0.20) & 7.74(0.01) & 0.46(0.02)     &  7.68 & \nodata \\
G180.92+04.53C1  & 0.96 & 9.79(0.59) & 1.53(0.03) & 3.71(0.11) & 3.49(0.11) & 1.43(0.02) & 2.13(0.05) & 0.85(0.11) & 1.27(0.07) & 1.63(0.16)        &  2.19 & RA \\
G180.92+04.53C2  & \nodata & 5.45(0.12) & 2.05(0.03) & 4.23(0.07) & 3.03(0.14) & 1.62(0.02) & 1.74(0.06) & 0.75(0.13) & 1.55(0.11) & 1.79(0.22)     &  2.20 & RA \\
\enddata
\tablenotetext{a}{The abbreviations are: RA: red asymmetry; BA: blue asymmetry; RP: red profile; BP: blue profile; wings: wings.}
\tablecomments{The complete table is available in machine-readable form.}
\end{deluxetable*}

\setcounter{table}{1}
\begin{deluxetable*}{ccccc ccccc c}
	\renewcommand{\thetable}{\arabic{table}B}
	\tablecaption{Statistics of observed parameters\label{stat-obs-para}}
	\tablehead{
	\nocolhead{} & \colhead{D} & \colhead{ T$_b$($^{12}$CO)} &
	\colhead{V$_{lsr}$($^{12}$CO)} & \colhead{${\rm \Delta V_{^{12}CO}}$} &
	\colhead{T$_b$($^{13}$CO)} & \colhead{ V$_{lsr}$($^{13}$CO)} &
	\colhead{${\rm \Delta V_{^{13}CO}}$} & \colhead{T$_b$(C$^{18}$O)} &
	\colhead{V$_{lsr}$(C$^{18}$O)} & \colhead{${\rm \Delta V_{C^{18}O}}$}  \\
	\nocolhead{} & \colhead{kpc} & \colhead{$K$} & \colhead{(km\ s$^{-1}$)} & \colhead{(km\ s$^{-1}$)} &
	\colhead{$K$} & \colhead{(km\ s$^{-1}$)} & \colhead{(km\ s$^{-1}$)} &
	\colhead{$K$} & \colhead{(km\ s$^{-1}$)} & \colhead{(km\ s$^{-1}$)}
	}
	\decimalcolnumbers
	\startdata
	\multicolumn{11}{c}{The First Galactic Quadrant (IQuad)}\\
	\hline
	Min. & 0.23 & 4.36 & -4.55 & 0.61 & 1 & -4.81 & 0.53 & 0.36 & -5.3 & 0.17           \\
	Max. & 4.15 & 19.01 & 29.2 & 6.47 & 9.36 & 29.07 & 4.39 & 3.34 & 29.01 & 3.98       \\
	Mean & 0.97  & 9.05  & 6.93  & 2.20  & 4.08  & 6.88  & 1.55  & 1.07  & 6.64  & 1.05 \\
	Median & 0.77 & 8.11 & 7.05 & 2.11 & 4.09 & 6.68 & 1.39 & 0.96 & 6.70 & 0.84       \\
	Standard deviation & 0.75  & 3.27  & 6.66  & 1.09  & 1.33  & 6.62  & 0.78  & 0.57  & 6.65  & 0.66 \\
	Median absolute deviation & 0.25 & 1.76 & 4.33 & 0.64 & 0.77 & 4.45 & 0.42 & 0.19 & 4.49 & 0.19 \\
	\hline
	\multicolumn{11}{c}{The Galactic Anti-Center Direction Region (ACent)}\\
	\hline
	Min. & 0.4 & 3.98 & -7.48 & 0.65 & 0.68 & -7.24 & 0.44 & 0.32 & -5.56 & 0.29  \\
	Max. & 9.17 & 18.31 & 46.17 & 5.65 & 7.1 & 46.28 & 5.15 & 2.9 & 19.46 & 2.28   \\
	Mean  & 1.46  & 8.67  & 7.29  & 2.71  & 3.67  & 7.28  & 1.77  & 0.89  & 4.61  & 1.07 \\
	Median & 1.54 & 8.10 & 5.27 & 2.41 & 3.53 & 5.28 & 1.54 & 0.75 & 4.67 & 0.99  \\
	Standard deviation & 1.54  & 2.93  & 8.94  & 1.16  & 1.48  & 8.95  & 0.87  & 0.49  & 5.13  & 0.51 \\
	Median absolute deviation & 0.65 & 1.98 & 3.61 & 0.76 & 0.97 & 3.75 & 0.42 & 0.03 & 1.36 & 0.05 \\
	\enddata
\end{deluxetable*}
	
\begin{deluxetable*}{ccccc cc}
\renewcommand{\thetable}{\arabic{table}A}
\tablecaption{Derived parameters at the core center\label{deri-para}}
\tablewidth{0pt}
\tablehead{
\colhead{source} & \colhead{T$_{\mathrm{ex}}$} & \colhead{$\tau_{13}$} &  \colhead{N$_{\mathrm{H}_2}$} &
\colhead{$\sigma_{\mathrm{NT}}$} & \colhead{$\sigma_{\mathrm{Th}}$} & \colhead{$\sigma_{3\mathrm{D}}$}  \\
\colhead{} & \colhead{(K)} & \colhead{} &  \colhead{(10$^{21}$ cm$^{-1}$)} &
\colhead{(km\ s$^{-1}$)} & \colhead{(km\ s$^{-1}$)} & \colhead{(km\ s$^{-1}$)}
}
\decimalcolnumbers
\startdata
\multicolumn{7}{c}{The First Galactic Quadrant (IQuad)}\\
\hline
G013.86+04.55C1    &   10.34(1.02)   &   0.41(0.04)   &   4.55 (0.94)   &   0.77(0.02)   &   0.18(0.01)   &   1.37(0.04)  \\
G013.86+04.55C2    &   11.58(1.02)   &   0.79(0.06)   &   8.31 (1.43)   &   0.60(0.01)   &   0.19(0.01)   &   1.09(0.02)  \\
G013.86+04.55C3    &   10.71(1.02)   &   0.93(0.07)   &   8.29 (1.50)   &   0.59(0.01)   &   0.18(0.01)   &   1.07(0.02)  \\
G013.86+04.55C4    &   11.16(1.02)   &   0.69(0.06)   &   8.07 (1.48)   &   0.72(0.01)   &   0.18(0.01)   &   1.29(0.02)  \\
G013.86+04.55C5    &   11.70(1.02)   &   0.74(0.08)   &   8.50 (1.61)   &   0.65(0.01)   &   0.19(0.01)   &   1.17(0.02)  \\
G014.74+04.06C1    &   9.72 (1.03)   &   0.68(0.05)   &   6.05 (1.18)   &   0.69(0.01)   &   0.17(0.01)   &   1.23(0.02)  \\
G014.74+04.06C2    &   14.32(1.01)   &   0.56(0.02)   &   9.04 (1.19)   &   0.64(0.00)   &   0.21(0.01)   &   1.17(0.01)  \\
G026.45+08.02C1    &   11.47(1.02)   &   1.13(0.19)   &   9.33 (2.14)   &   0.48(0.01)   &   0.19(0.01)   &   0.89(0.02)  \\
G028.45-06.39C1    &   11.42(1.02)   &   1.06(0.06)   &   9.24 (1.51)   &   0.51(0.00)   &   0.19(0.01)   &   0.94(0.01)  \\
G028.45-06.39C2    &   11.43(1.02)   &   1.06(0.05)   &   8.46 (1.37)   &   0.47(0.00)   &   0.19(0.01)   &   0.88(0.01)  \\
\hline
\multicolumn{7}{c}{The Galactic Anti-Center Direction Region (ACent)}\\
\hline
G178.72-07.01C1    &   13.29(1.01)   &   1.20(0.11)   &   7.72 (1.28)   &   0.29(0.00)   &   0.20(0.01)   &   0.61(0.01)  \\
G178.98-06.74C1    &   11.39(1.02)   &   1.28(0.14)   &   8.61 (1.64)   &   0.40(0.01)   &   0.19(0.01)   &   0.77(0.02)  \\
G178.98-06.74C2    &   11.87(1.02)   &   1.33(0.12)   &   8.95 (1.57)   &   0.37(0.00)   &   0.19(0.01)   &   0.72(0.01)  \\
G178.98-06.74C3    &   12.15(1.02)   &   1.54(0.13)   &   9.69 (1.61)   &   0.33(0.00)   &   0.19(0.01)   &   0.66(0.01)  \\
G180.92+04.53C1    &   13.21(1.01)   &   0.44(0.04)   &   8.66 (1.39)   &   0.90(0.01)   &   0.20(0.01)   &   1.60(0.02)  \\
G180.92+04.53C2    &   8.77 (1.03)   &   0.81(0.06)   &   6.50 (1.37)   &   0.74(0.01)   &   0.16(0.01)   &   1.31(0.03)  \\
G181.42-03.73C1    &   11.70(1.02)   &   0.65(0.05)   &   7.04 (1.20)   &   0.61(0.01)   &   0.19(0.01)   &   1.11(0.02)  \\
G181.42-03.73C2    &   11.69(1.02)   &   0.93(0.05)   &   8.98 (1.45)   &   0.55(0.01)   &   0.19(0.01)   &   1.01(0.02)  \\
G181.42-03.73C3    &   13.68(1.01)   &   0.80(0.04)   &   9.79 (1.41)   &   0.52(0.01)   &   0.20(0.01)   &   0.96(0.02)  \\
G181.42-03.73C4    &   12.75(1.02)   &   0.83(0.05)   &   8.02 (1.22)   &   0.47(0.00)   &   0.20(0.01)   &   0.88(0.01)  \\
\enddata
\tablecomments{The complete table is available in machine-readable form.}
\end{deluxetable*}

\setcounter{table}{2}
\begin{deluxetable*}{ccccc cc}
	\renewcommand{\thetable}{\arabic{table}B}
	\tablecaption{Statistics of derived parameters
	\label{stat-deri-para}}
	\tabletypesize{\scriptsize}
	\tablehead{
	\nocolhead{}  & \colhead{T$_{\mathrm{ex}}$} & \colhead{$\tau_{13}$}  & \colhead{N$_{\mathrm{H}_2}$} &
	\colhead{$\sigma_{\mathrm{NT}}$} & \colhead{$\sigma_{\mathrm{Th}}$} & \colhead{$\sigma_{3\mathrm{D}}$}\\
	\nocolhead{}  & \colhead{(K)} & \colhead{}  & \colhead{($10^{21}$ $cm^{-1}$)} &
	\colhead{(km\ s$^{-1}$)} & \colhead{(km\ s$^{-1}$)} & \colhead{(km\ s$^{-1}$)}
	}
	\decimalcolnumbers
	\startdata
	\multicolumn{7}{c}{The First Galactic Quadrant (IQuad)}\\
	\hline
	Min. & 7.64 & 0.13 & 0.68 & 0.22 & 0.15 & 0.48\\
	Max. & 22.5 & 3.78 & 68.67 & 1.86 & 0.26 & 3.24\\
	Mean & 12.44  & 0.79  & 9.11  & 0.65  & 0.19  & 1.19\\
	Median & 11.50  & 0.63  & 7.23  & 0.59  & 0.19  & 1.07\\
	Standard deviation & 3.32  & 0.56  & 8.69  & 0.33  & 0.03  & 0.56 \\
	Median absolute deviation & 1.79 & 0.27 & 2.43 & 0.18 & 0.02 & 0.30 \\
	\hline
	\multicolumn{7}{c}{The Galactic Anti-Center Direction Region (ACent)}\\
	\hline
	Min. & 7.24 & 0.07 & 0.56 & 0.18 & 0.15 & 0.47      \\
	Max. & 21.8 & 2.43 & 27.33 & 2.19 & 0.26 & 3.82     \\
	Mean & 12.05  & 0.64  & 8.15  & 0.75  & 0.19  & 1.35\\
	Median & 11.49  & 0.58  & 7.50  & 0.65  & 0.19  & 1.18\\
	Standard deviation & 2.98  & 0.36  & 5.46  & 0.37  & 0.02  & 0.63 \\
	Median absolute deviation & 2.03 & 0.22 & 2.65 & 0.18 & 0.02 & 0.30 \\
\enddata
\end{deluxetable*}
	
\begin{deluxetable*}{ccccc cccc}
\renewcommand{\thetable}{\arabic{table}A}
\tablecaption{Emission region parameters\label{area-para}}
\tablewidth{0pt}
\tablehead{
\colhead{source} & \colhead{offset} &
\colhead{$a\times b$\tablenotemark{\,{\footnotesize 1}}} & \colhead{R} &  \colhead{n$_{vol}$}  &
\colhead{M} & \colhead{M$_{\mathrm{vir}}$}  &
\colhead{M$_{\mathrm{Jeans}}$} & \colhead{$\alpha_{\mathrm{vir}}$}\\
\colhead{} & \colhead{'',''} &
\colhead{$''\times ''$} & \colhead{(pc)} &  \colhead{(10$^3$ cm$^{-3}$)} &
\colhead{(M$_{\bigodot}$)} & \colhead{(M$_{\bigodot}$)}  &
\colhead{(M$_{\bigodot}$)} & \colhead{}
}
\decimalcolnumbers
\startdata
\multicolumn{9}{c}{The First Galactic Quadrant (IQuad)}\\
\hline
G013.86+04.55C1   &   (-121,292)   &    $197\times 69$    &   0.08   &  8.97    &   1     &  31    &  92   &  21.74\\
G013.86+04.55C2   &   (97,84)      &    $201\times 120$   &   0.11   &  12.33   &   5     &  25    &  39   &  5.59 \\
G013.86+04.55C3   &   (-48,86)     &    $161\times 155$   &   0.11   &  12.04   &   5     &  24    &  38   &  5.19 \\
G013.86+04.55C4   &   (-241,95)    &    $252\times 188$   &   0.15   &  8.53    &   9     &  50    &  78   &  5.73 \\
G013.86+04.55C5   &   (102,-175)   &    $197\times 149$   &   0.12   &  11.39   &   6     &  33    &  51   &  5.73 \\
G014.74+04.06C1   &   (2,140)      &    $196\times 133$   &   0.42   &  2.31    &   50    &  129   &  131  &  2.60 \\
G014.74+04.06C2   &   (-58,-95)    &    $236\times 212$   &   0.59   &  2.49    &   143   &  152   &  107  &  1.07 \\
G026.45+08.02C1   &   (263,148)    &    $247\times 197$   &   0.13   &  11.75   &   7     &  19    &  22   &  2.69 \\
G028.45-06.39C1   &   (119,-52)    &    $179\times 117$   &   0.08   &  17.69   &   3     &  14    &  21   &  4.73 \\
G028.45-06.39C2   &   (7,-26)      &    $142\times 121$   &   0.08   &  17.90   &   2     &  11    &  17   &  4.72 \\
\hline
\multicolumn{9}{c}{The Galactic Anti-Center Direction Region (ACent)}\\
\hline
G178.72-07.01C1   &   (-247,-22)   &    $243\times 182$   &   0.32   &  3.88    &   37    &  17    &  12   &  0.48 \\
G178.98-06.74C1   &   (-170,267)   &    $161\times 124$   &   0.20   &  6.90    &   16    &  20    &  18   &  1.26 \\
G178.98-06.74C2   &   (-110,109)   &    $137\times 89$    &   0.16   &  9.15    &   10    &  14    &  13   &  1.36 \\
G178.98-06.74C3   &   (19,-62)     &    $210\times 141$   &   0.25   &  6.37    &   27    &  17    &  12   &  0.65 \\
G180.92+04.53C1   &   (-41,8)      &    $199\times 124$   &   0.37   &  3.82    &   53    &  189   &  222  &  3.54 \\
G180.92+04.53C2   &   (-104,-69)   &    $110\times 87$    &   0.23   &  4.63    &   15    &  78    &  112  &  5.09 \\
G181.42-03.73C1   &   (348,31)     &    $171\times 118$   &   0.60   &  1.88    &   118   &  142   &  105  &  1.21 \\
G181.42-03.73C2   &   (-95,-12)    &    $174\times 121$   &   0.61   &  2.36    &   155   &  118   &  71   &  0.76 \\
G181.42-03.73C3   &   (-108,-139)  &    $199\times 125$   &   0.67   &  2.37    &   201   &  117   &  62   &  0.58 \\
G181.42-03.73C4   &   (-91,-248)   &    $101\times 92$    &   0.41   &  3.17    &   61    &  58    &  42   &  0.95 \\
\enddata
\tablenotetext{1}{The a and b are Gaussian semi-major axis and semi-minor axis.}
\tablecomments{The complete table is available in machine-readable form.}
\end{deluxetable*}

\setcounter{table}{3}
\begin{deluxetable*}{ccccc c}
	\renewcommand{\thetable}{\arabic{table}B}
	\tablecaption{Statistics of region parameters.
	\label{stat-area-para}}
	\tabletypesize{\scriptsize}
	\tablehead{
	\nocolhead{}  & \colhead{R} & \colhead{n$_{vol}$} & \colhead{M} & \colhead{M$_{\mathrm{vir}}$}  &
	\colhead{M$_{\mathrm{Jeans}}$} \\
	\nocolhead{} & \colhead{(pc)} & \colhead{(10$^3$ cm$^{-3}$)} & \colhead{(M$_{\bigodot}$)} & \colhead{(M$_{\bigodot}$)}  &
	\colhead{(M$_{\bigodot}$)}
	}
	\decimalcolnumbers
	\startdata
	\multicolumn{6}{c}{The First Galactic Quadrant (IQuad)}\\
	\hline
	Min. &    0.04  &  0.34  &  0.1  &  3    &  4 \\
	Max. &    2.94  &  17.90  &  7613  &  3300  &  2357 \\
	Mean &   0.44  &  4.42  &  291  &  248  &  186 \\
	Median &   0.32  &  3.23  &  31   &  72   &  70 \\
	Standard deviation  &  0.41  &  3.28  &  937  &  534  &  348 \\
	Median absolute deviation & 0.14 & 1.38 & 26 & 57 & 53 \\
	\hline
	\multicolumn{6}{c}{The Galactic Anti-Center Direction Region (ACent)}\\
	\hline
	Min. &  0.09  &  0.43  &  1    &  5    &  8   \\
	Max. &    2.82  &  21.65  &  3179  &  1504  &  3121 \\
	Mean &   0.47  &  5.22  &  151  &  211  &  244   \\
	Median &  0.43  &  2.63  &  43   &  110  &  106  \\
	Standard deviation &  0.45  &  5.53  &  449  &  301  &  410  \\
	Median absolute deviation & 0.21 & 1.71 & 34 & 79 & 68 \\
	\enddata
\end{deluxetable*}

\begin{deluxetable*}{ccccccc| cccc}
	\tablecaption{Statistics of $\Delta V_{\rm ^{13}CO}$, $\sigma_{3\mathrm{D}}$, T$_{ex}$ , N$_{\rm H_2}$ and M in different sub-regions
	\label{stat-complexes}}
	\tabletypesize{\scriptsize}
	\tablehead{
	\nocolhead{} & \colhead{Serpens} & \colhead{Aquila} & \colhead{Cloud A/B} & \colhead{Vulpecula} & \colhead{Cygnus} & \colhead{Cepheus} & \colhead{Auriga} & \colhead{Gemini} & \colhead{Monoceros} & \colhead{Canis Major}
	}
	\decimalcolnumbers
	\startdata
\hline
all cores\tablenotemark{\,{\footnotesize a}} & 7 & 19 & 19 & 27 & 67 & 7 & 26 & 20 & 40 & 11 \\
\hline
regions & \multicolumn{6}{c}{IQuad} \vline& \multicolumn{4}{c}{ACent} \\
\hline
\multicolumn{11}{c}{$\Delta V_{\rm ^{13}CO}$ (km s$^{-1}$)} \\
\hline
min & 1.39  & 0.56  & 0.53  & 0.78  & 0.72  & 1.08  & 0.44  & 0.72  & 0.75  & 0.80                        \\
max & 1.81  & 2.03  & 1.85  & 2.54  & 4.39  & 1.45  & 2.31  & 3.01  & 5.15  & 3.15                        \\
mean & 1.57  & 1.05  & 0.88  & 1.47  & 1.93  & 1.26  & 1.41  & 1.80  & 1.90  & 2.24                       \\
median & 1.54  & 0.91  & 0.67  & 1.54  & 1.71  & 1.28  & 1.37  & 1.58  & 1.48  & 2.26                     \\
std & 0.14  & 0.44  & 0.42  & 0.54  & 0.87  & 0.15  & 0.52  & 0.69  & 1.08  & 0.67                        \\
\hline
\multicolumn{11}{c}{$\sigma_{3\mathrm{D}}$ (km s$^{-1}$)}      \\
\hline
min & 1.07  & 0.49  & 0.48  & 0.66  & 0.59  & 0.84  & 0.47  & 0.61  & 0.60  & 0.66    \\
max & 1.37  & 1.52  & 1.38  & 1.90  & 3.24  & 1.10  & 1.74  & 2.25  & 3.82  & 2.35    \\
mean & 1.20  & 0.83  & 0.72  & 1.14  & 1.47  & 0.97  & 1.09  & 1.36  & 1.44  & 1.68   \\
median & 1.17  & 0.72  & 0.59  & 1.16  & 1.30  & 0.99  & 1.05  & 1.20  & 1.12  & 1.70 \\
std & 0.10  & 0.31  & 0.29  & 0.38  & 0.63  & 0.11  & 0.36  & 0.49  & 0.78  & 0.48    \\
\hline
\multicolumn{11}{c}{T$_{\mathrm{ex}}$ (K)}      \\
\hline
min & 9.72  & 8.45  & 7.64  & 7.72  & 7.71  & 8.41  & 7.47  & 7.59  & 7.24  & 9.42                        \\
max & 14.32  & 20.74  & 13.80  & 22.50  & 22.43  & 12.85  & 16.94  & 16.70  & 21.80  & 14.81              \\
mean & 11.36  & 11.11  & 10.94  & 13.44  & 13.13  & 10.67  & 11.73  & 11.45  & 12.57  & 12.14             \\
median & 11.16  & 10.64  & 11.32  & 12.92  & 12.03  & 10.66  & 11.87  & 10.68  & 11.33  & 13.33           \\
std & 1.37  & 2.66  & 1.51  & 3.73  & 3.59  & 1.40  & 2.17  & 2.43  & 3.76  & 2.06                        \\
\hline
\multicolumn{11}{c}{N$_{\mathrm{H_2}}$ (10$^{21}$ cm$^{-2}$)}      \\
\hline
min & 4.55  & 1.34  & 2.16  & 1.86  & 0.68  & 4.81  & 0.56  & 2.62  & 1.55  & 2.95                        \\
max & 9.04  & 18.66  & 10.45  & 18.42  & 68.67  & 9.23  & 12.18  & 22.73  & 27.33  & 19.46                \\
mean & 7.54  & 6.83  & 4.93  & 7.26  & 12.06  & 7.12  & 7.37  & 7.63  & 8.48  & 9.97                      \\
median & 8.29  & 7.21  & 3.88  & 6.65  & 7.88  & 7.53  & 8.02  & 5.13  & 6.62  & 7.71                     \\
std & 1.50  & 4.03  & 2.55  & 3.68  & 11.62  & 1.62  & 2.57  & 5.25  & 6.74  & 5.69                       \\
\hline
\multicolumn{11}{c}{M (M$_{\bigodot}$)}      \\
\hline
min & 1.4 & 0.1 & 2.8 & 1.3 & 1 & 37.6 & 1.2 & 15.2 & 0.7 & 2.7  \\
max & 142.5 & 15.4 & 36.5 & 2367.7 & 7612.8 & 147.7 & 244.6 & 338.5 & 840.5 & 18.9   \\
mean & 31.0  & 4.1  & 13.6  & 214.0  & 515.6  & 71.1  & 77.4  & 108.9  & 76.1  & 9.0 \\
median & 5.7 & 2.35 & 11.2 & 36.4 & 62.8 & 54.1 & 59.55 & 72.8 & 37.6 & 5.9    \\
std & 48.1  & 4.4  & 9.9  & 486.4  & 1302.7  & 38.5  & 69.4  & 92.4  & 144.8  & 5.7  \\
\enddata
\tablenotetext{a}{The countings of cores in different sub-regions.}
\end{deluxetable*}

\begin{deluxetable*}{cccc}
	\tablecaption{Fifty-eight cores associated with objects
	\label{association}
	}
	\tabletypesize{\scriptsize}
	\tablehead{
	\colhead{cores} &  \colhead{Class I/II YSO candidates} & \colhead{Class III YSO candidates} &\colhead{IRAS point source}
	}
	\decimalcolnumbers
\startdata
\multicolumn{4}{c}{The First Galactic Quadrant (IQuad)}      \\
\hline
G013.86+04.55C1  &  0  &  1  &  0  \\
G013.86+04.55C4  &  0  &  1  &  0  \\
G013.86+04.55C5  &  0  &  1  &  0  \\
G015.16+07.23aC1  &  0  &  0 & 1 \\
G015.16+07.23bC5  &  0  &  0 & 1 \\
G028.45-06.39C2  &  0  &  0 & 1 \\
G034.69-06.57C4  &  0  &  1  &  0  \\
G038.36-00.95bC1  &  0  &  0 & 1 \\
G048.40-05.82C1  &  0  &  0 & 1 \\
G049.06-04.18C1  &  0  &  1  &  1  \\
G052.99+03.07bC2  &  0  &  0 & 1 \\
G057.17+03.41C3  &  0  &  1  &  0  \\
G057.26+04.01C1  &  0  &  0 & 1 \\
G058.16+03.50C2  &  1  &  0  &  1  \\
G058.97-01.66bC1  &  0  &  0 & 1 \\
G060.75-01.23C2  &  1  &  0  &  1  \\
G065.30-08.44C1  &  0  &  1  &  0  \\
G070.44-01.54C5  &  0  &  0 & 1 \\
G084.79-01.11dC1 &  1  &  0  &  0  \\
G089.29+04.01C2  &  1  &  0  &  1  \\
G089.36-00.67C1  &  2  &  0  &  1  \\
G089.64-06.59C1  &  1  &  0  &  1  \\
G089.75-02.16C3  &  0  &  0 & 1 \\
G089.93-07.03C1  &  1  &  0 & 1  \\
G090.76-04.55C1  &  0  &  0 & 1 \\
G092.79+09.14C3  &  0  &  0 & 1 \\
G093.75-04.59C1  &  1  &  0 & 1  \\
G093.99-04.91C3  &  1  &  0  &  0  \\
G093.91+10.02C2  &  1  &  0 & 1  \\
G095.51+09.97aC1  &  0  &  0 & 1  \\
G095.51+09.97bC1 &  1  &  0  &  0  \\
G097.20+09.87C2  &  0  &  0 & 1  \\
G097.77+08.59C1  &  3  &  0  &  1  \\
G097.77+08.59C3  &  3  &  0  &  0  \\
\hline
All  & 18  &  7  & 24  \\
\hline
\multicolumn{4}{c}{The Galactic Anti-Center Direction region (ACent)}      \\
\hline
G180.92+04.53C1  &  2  &  0  &  0  \\
G180.92+04.53C2  &  1  &  0  &  0  \\
G181.42-03.73C1  &  0  &  0 & 1  \\
G181.42-03.73C2  &  0  &  0 & 1  \\
G181.42-03.73C3  &  1  &  0 & 1  \\
G181.84+00.31bC1 &  3  &  0 & 1   \\
G181.84+00.31bC4 &  1  &  0 & 0   \\
G182.04+00.41bC1 &  1  &  0 & 1   \\
G185.33-02.12C2  &  2  &  0 & 1   \\
G185.33-02.12C3  &  0  &  0 & 1  \\
G188.04-03.71C3  &  1  &  0 & 1   \\
G201.13+00.31C2  &  0  &  0 & 1  \\
G201.26+00.46bC2 &  1  &  0  &  0  \\
G201.44+00.65C2  &  1  &  0  &  0  \\
G201.59+00.55C2  &  1  &  0  &  0  \\
G201.84+02.81C1  &  1  &  0  &  0  \\
G202.58-08.70C1  &  0  &  0 & 1  \\
G203.00-03.69C2  &  1  &  0  &  0  \\
G212.18+05.11C2  &  0  &  0 & 1  \\
G224.47-00.65C1  &  2  &  0 & 1  \\
G226.29-00.63C3  &  2  &  1 & 1  \\
G226.36-00.50C1  &  3  &  0  &  1  \\
G226.36-00.50C2  &  4  &  1 & 1  \\
G226.80-07.04C1  &  1  &  0 & 1  \\
\hline
All &  29  & 2  & 16 \\
\enddata
\end{deluxetable*}

\end{document}